\documentclass[twocolumn]{aastex631}

\usepackage{graphicx}%
\usepackage{amsmath,amssymb,amsfonts}%
\usepackage{amsthm}%
\usepackage{mathrsfs}%
\usepackage{tablefootnote}
\usepackage{booktabs}%
\usepackage{threeparttable}

\newcommand{\crit}{\text{crit}}

\begin{document}

%\title{From QPEs to bright sirens: GW detectability and $H_0$ measurement under secular evolution}
 \title{Hubble constant measurement from QPEs as electromagnetic counterparts to extreme mass ratio inspirals}

%Precise Hubble constant measurement from quasi-periodic eruptions as electromagnetic counterparts to extreme mass ratio inspirals  
\correspondingauthor{Fa-Yin Wang}
\email{fayinwang@nju.edu.cn}

\author[0009-0007-6439-6891]{Yejing Zhan}
\affiliation{School of Astronomy and Space Science, Nanjing University, Nanjing 210093, People’s Republic of China;}

\author[0000-0003-1357-7135]{Di Wang}
\affiliation{School of Astronomy and Space Science, Nanjing University, Nanjing 210093, People’s Republic of China;}

\author[0000-0003-0672-5646]{Shuang-Xi Yi}
\affiliation{School of Physics and Physical Engineering, Qufu Normal University, Qufu 273165, People’s Republic of China;}
\author[0000-0003-4157-7714]{Fa-Yin Wang}
\affiliation{School of Astronomy and Space Science, Nanjing University, Nanjing 210093, People’s Republic of China;}
\affiliation{Key Laboratory of Modern Astronomy and Astrophysics (Nanjing University), Ministry of Education, Nanjing 210093, People's Republic of China}
% 


% \affil[2]{\orgdiv{School of Physics and Physical Engineering}, \orgname{Qufu Normal University}, \orgaddress{\city{Qufu}, \postcode{273165}, \country{China}}}
\begin{abstract}
    Gravitational waves (GWs) accompanied by electromagnetic (EM) counterparts provide a novel methodology to measure the Hubble constant ($H_0$), known as bright sirens. However, the rarity of such multi-messenger events limits the precision of the $H_0$ constraint. Recently, the newly-discovered nuclear transient, quasi-periodic eruptions (QPEs) show intriguing evidence of a stellar-mass companion captured by a supermassive black hole (SMBH) in an extreme/intermediate mass-ratio inspiral (EMRI/IMRI), which is the most promising sources of the space-based GW detectors, such as LISA. 
    % Here, we study the secular orbital evolution of QPE systems under different theoretical frameworks based on timing observation, including the stripping scenario and the orbiter-disk scenario, and assess their GW detectability by LISA.
    Here, we model the secular orbital evolution of known QPE systems using two frameworks: a stripping scenario in which periodic mass transfer at periapsis drives the evolution; and an orbiter–disk collision scenario in which the companion interacts with a misaligned accretion disk, modulated by coupled orbiter-disk precession. For each framework, we assess detectability by LISA, together with the resulting constraints on $H_0$. Our principal findings are: (i) in the stripping scenario, no currently known QPE reaches detectability within a four-year LISA mission. (ii) in the orbiter–disk scenario, two sources—eRO-QPE2 and eRO-QPE4—are detectable with signal-to-noise ratios $\simeq 8.5-28.8$ and constrain $H_0$ with fractional uncertainty of 6.7-14.9\%. QPE systems remain uncertain on the decade-long secular evolution. Therefore, they motivate continued time-domain monitoring of QPE candidates.
\end{abstract}

\section{introduction}

%%% 核心点是给出了bright siren的新模型，并用来计算了对Hubble constant 的限制
\par The precise measurement of Hubble constant $(H_0)$ is a crucial task for validating the standard cosmological model, as a significant tension has been found between measurements from the late and early universe \citep{verdeTensionsEarlyLate2019,aghanimPlanck2018Results2020,riessComprehensiveMeasurementLocal2022,huHubbleTensionEvidence2023}. The cosmic microwave background from the early universe provides a model-dependent constraint on $H_0$, relying on the assumption of the standard cosmological model. In contrast, $H_0$ can be measured by the local distance ladder, using a well-approximated relation
\begin{equation}
    v_r = H_0d_L,
    \label{eq:hubble}
\end{equation}
where $v_r$ is the recessional velocity, and $d_L$ is the luminosity distance.
However, this method is prone to error accumulation during the calibration of the distance ladder. The bright siren method offers an independent alternative, providing direct distance measurement from gravitational waves (GWs), thereby enabling a model-independent measurement of $H_0$ \citep{schutzDeterminingHubbleConstant1986,holzUsingGravitationalWaveStandard2005,abbottGravitationalwaveStandardSiren2017}. Following the first detection of the bright siren GW170817/GRB 170817A in 2017 \citep{ligoscientificcollaborationandvirgocollaborationGW170817ObservationGravitational2017}, there have been no more GW signals of bright sirens detected by the current ground-based GW detectors. The challenges associated with detecting and localizing GW sources using ground-based detectors are significant \citep{groverComparisonGravitationalWave2014,chassande-mottinGravitationalWaveObservations2019}. These detectors suffer from poor sky angle localization compared to those in space, leading to a large number of candidate host galaxies when searching for electromagnetic (EM) counterparts. Besides, the primary targets for bright sirens from ground-based observatories are neutron star-neutron star (NS-NS) mergers and neutron star-black hole (NS-BH) mergers, which are associated with EM counterparts such as gamma-ray bursts (GRBs) and kilonovae. The latest O4 run of the ground-based observatory network, LIGO-Virgo-Kagra, indicates that the event rate of these mergers is only $\sim 0.2-5 \text{ yr}^{-1}$ \citep{collaborationGWTC40UpdatingGravitationalWave2025}, under the detection horizon of the O4 extending to $\sim 160 \text{ Mpc}$ \citep{collaborationGWTC40IntroductionVersion2025}. 
% The event rate of these mergers is only $\mathcal{O}(0.1) \text{ yr}^{-1}$ 
%\citep{mapelliCosmicMergerRate2018,chruslinskaDoubleNeutronStars2018,abbottObservationGravitationalWaves2021} under the detection horizon of the current ground-based detectors extending to $\sim 0.1\text{ Gpc}$ \citep{chenDistanceMeasuresGravitationalwave2021,ligoscientificcollaborationandvirgocollaborationGW170817ObservationGravitational2017}. 
Compounding this issue, only about $30\%$ of kilonovae associated with the mergers are detectable \citep{metzgerKilonovae2019}. The detection rate of corresponding GRBs is even lower, due to the beaming effect \citep{wandermanRateLuminosityFunction2015,Zhang2018}. These challenges make it difficult to detect a sufficient number of NS-NS and NS-BH bright sirens. These limitations collectively hinder the critical need to identify novel classes of bright sirens.  
%Thus, the EMRI bright siren is a more promising standard siren in the next decade. %The dilemma of NS-NS bright siren is expected to be solved in the coming few decades with the deployment of the next-generation ground-based detectors \citep{Abac2025}, which would extend the detection horizon for NS-NS and NS-BH mergers to $o(10) \text{Gpc}$.

\par Millihertz GW sources, known as extreme mass-ratio inspirals (EMRIs), and their possible EM counterparts, quasi-periodic eruptions (QPEs), circumvent these challenges. The EM counterparts offer a high event rate \citep{arcodiaCosmicHideSeek2024}, and are sufficiently bright for detection. Besides, space-based GW detectors, like LISA, Taiji, and Tianqin \citep{amaro-seoaneAstrophysicsLaserInterferometer2023,luoTianQinSpaceborneGravitational2016,huTaijiProgramSpace2017}, will observe EMRIs with a more precise localization, enabling unambiguous EM counterpart identification. 
%Here, we explore the expected constraint on $H_0$ from EMRIs with QPEs as EM counterparts. 

% QPE
%\par QPEs are X-ray bursts that flare up periodically, on a time scale of hours, from the nuclei of galaxies. 
\par QPE represents a class of X-ray bursts characterized by their periodic flaring activity, which occurs on a time scale of hours and emanates from the centers of galaxies \citep{miniuttiNinehourXrayQuasiperiodic2019,nichollOutflowPowersOptical2020,arcodiaXrayQuasiperiodicEruptions2021,giustiniXrayQuasiperiodicEruptions2020,arcodiaMoreMerrierSRG2024,nichollQuasiperiodicXrayEruptions2024,chakrabortyDiscoveryQuasiperiodicEruptions2025,hernandez-garciaDiscoveryExtremeQuasiPeriodic2025}. 
Up to the time of writing this paper, nine QPE sources have been discovered. Their observational characteristics are presented in Table \ref{tab:sources}. Despite extensive study, the physical origin of QPEs continues to be a subject of debate. The current leading theoretical model is that QPEs involve a stellar-mass companion orbiting a supermassive black hole (SMBH).
This type of model is categorized into two primary scenarios. One is the stripping scenario, in which QPE is produced by mass transfer from a compact companion at pericenter, such as a white dwarf or an evolved star \citep{kingGSN069Tidal2020,zhaoQuasiperiodicEruptionsHelium2022,wangModelPossibleConnection2022,luQuasiperiodicEruptionsMildly2023,krolikQuasiperiodicEruptersStellar2022,linialUnstableMassTransfer2023}. The other is the orbiter-disk collision scenario, which suggests that a QPE is produced by the collision between a stellar or intermediate mass object and an accretion disk surrounding a SMBH \citep{xianXRayQuasiperiodicEruptions2021,linialEMRITDEQPE2023,franchiniQuasiperiodicEruptionsImpacts2023,tagawaFlaresStarsCrossing2023,yaoStarDiskCollisionsImplications2024,linialQPEsEMRIDebris2025}. In both scenarios, QPEs are inevitable EM counterparts of EMRIs, which are the main sources of millihertz GWs expected for LISA and prospective bright sirens.

% \par Recently, several less intense eruptions around SMBHs, called quasi-periodic oscillations (QPOs), have been discovered \citep{gierlinskiPeriodicity1Hour2008,mastersonMillihertzOscillationsInnermost2025}. QPOs show some similarities to QPEs, such as flaring properties, the properties of the host galaxies, and SMBHs. But the periods of QPOs (from about 7 to 60 minutes) are much shorter than those of QPEs. These QPOs are also probably produced by extreme mass ratio companions around SMBHs \citep{mastersonMillihertzOscillationsInnermost2025}, which increases the number of EMRI bright sirens.   

%最新的QPE的要放在摘要里已知的QPE (checked)
% \clearpage
\begin{table*}[htbp]
\caption{Observed characteristics of QPEs}% potential bright sirens}
\label{tab:sources}

\hspace{-2cm}
\begin{threeparttable} 
\begin{tabular}{@{}l@{\hspace{1cm}}c@{\hspace{1cm}}ccc@{}}
\toprule
Name            & $z$   &\parbox{1.5cm}{\centering $\sigma_{v_r}$\tnote{*}} & \parbox{3cm}{\centering recurrent time\\ $(\text{hours})$}  & \parbox{3cm}{\centering SMBH mass\\ $(\times 10^{6}M_\odot)$}  \\ \midrule
GSN 069\citep{miniuttiNinehourXrayQuasiperiodic2019}      & 0.0181 & 63 & 9                                                                         & $0.3-3.1$                                                                                                                 \\
eRO-QPE1\citep{arcodiaXrayQuasiperiodicEruptions2021}      & 0.0505&56 & 19                                                                        & $0.2-2.1$                                                                                                                      \\
eRO-QPE2\citep{arcodiaXrayQuasiperiodicEruptions2021}       & 0.0175&36 & 2.4                                                                       & $0.03-0.3$                                                                                                                 \\
RX J1301.9+2747\citep{giustiniXrayQuasiperiodicEruptions2020}  & 0.0237 &90 & 4.5                                                                       & $0.8-2.8$                                                                                                                  \\
eRO-QPE3\citep{arcodiaMoreMerrierSRG2024}      & 0.024 &152 & 20                                                                        & $0.4-1.7$                                                                                                               \\
eRO-QPE4\citep{arcodiaMoreMerrierSRG2024}        & 0.044 &133 & 11.5                                                                      & $17-68$                                                                                                                    \\
AT2019qiz\citep{nichollOutflowPowersOptical2020}       & 0.0151 &- & 48                                                                      & $\sim 1$                                                                                                                     \\
AT2022upj\citep{chakrabortyDiscoveryQuasiperiodicEruptions2025}       & 0.054 &- & 12-84   & $0.6-1.6$             
\\ZTF19acnskyy\citep{hernandez-garciaDiscoveryExtremeQuasiPeriodic2025}       & 0.024  & 72 & $\sim 108$                                                                & $\sim 1$              
\\
% RE J1034+396       & 0.042 & 68 & 1                                                                     & $1-10$    & QPO      \\
% 1ES 1927+654   & 0.019& - & $<0.3$                      & $0.6-2.5$          & QPO\tnote{**} \\
\bottomrule
\end{tabular}

\begin{tablenotes}        
        \footnotesize
        \centering
        \item[*] The $\sigma_{v_r}$ data are taken from \cite{bianHostGalaxyNarrowline2010,weversHostGalaxyProperties2022,sanchez-saezSDSS1335+0728Awakening1062024}.
        % \item[**] If 1ES 1927+654 is an EMRI, the nature of its comparison is completely unstudied. We do not constrain $H_0$ using this source.  
      \end{tablenotes} 
\end{threeparttable} 
\end{table*}

\par Regardless of the specific models, the typical QPE recurrent time ($P\sim \text{hours} \sim \mathcal O(10^{4}) \text{ s}$) corresponds to the GW emission frequency of $\mathcal{O}(10^{-5}) \text{ Hz}$, which is far below the optimal sensitivity band of the space-based GW detectors ($\mathcal{O}(10^{-3}) \text{ Hz}$), indicating a null detectability of these QPE sources at present \citep{chenMilliHertzGravitationalwaveBackground2022}. However, observations show that the recurrence time $T_\text{rec}$ decays regularly at a fast rate about $\dot T_\text{rec} \sim \mathcal O(10^{-5}) \text{ s s}^{-1} $ in some QPEs, such as GSN 069 and eRO-QPE2 \citep{miniuttiEppurSiMuove2025,arcodiaTickingAwayLongterm2024}, suggesting a rapid orbital evolution $\dot P_\text{orb}$. Thus, some QPE sources may shift their GW emission into the detectable band of the space-based GW detectors within a yearly timescale ($P_\text{orb}/\dot P_\text{orb}\sim \mathcal O(10^{9}) \text{ s}\sim \mathcal O(30) \text{ years}$). In contrast, QPEs, like eRO-QPE1 and RX J1301.9+2747, exhibit irregular evolution, which current models struggle to explain, though the irregular periodicity may be explained by the migration trap effect generated by the torque forces of the EMRI environment \citep{bellovaryMIGRATIONTRAPSDISKS2016,kejriwalRepeatingNuclearTransients2024}.  Additionally, the recently identified sources eRO-QPE3, eRO-QPE4, AT2019qiz, AT2022upj, and ZTF19acnskyy lack sufficient long-term monitoring to resolve their recurrence dynamics.          

\par Notably, observations on eRO-QPE2 indicate a significant evolution on $\dot T_\text{rec}$. The recurrence time decay rate evolves from $\dot T_\text{rec}\sim \mathcal O(10^{-5}) \text{ s s}^{-1}$ in the period of 2020-2022 to $\dot T_\text{rec}\lesssim \mathcal O(10^{-6}) \text{ s s}^{-1}$ during 2022-2023 \citep{arcodiaTickingAwayLongterm2024}. This evolution provides an additional constraint on the QPE models and complicates the analysis of GW detectability.
%%%% 说我们要做什么， 考虑长期演化，尽可能符合观测的时变。强调我们不是做模型，只是分析GW detectability

% 这里要写QPE2展现出Pdot停止的现象。regular 和 irregular 都讨论，irregular 给negative的结果就可以了

\par In this work, we calculate the secular evolution of the known QPE sources in two scenarios (the stripping and the orbiter-disk scenarios) to assess their GW detectability in the 2030s. We limit our analysis to the systems with regular orbital evolution, and only discuss some of the other unresolved systems for reference. We summarize the calculation of orbital evolution for both scenarios and the method for parameter estimation in Sect. \ref{sec:Method}. The detectability for the known sources and their constraint on $H_0$ are presented in Sect \ref{sec:result}. Additionally, we discuss additional potential effects for GW detectability in Sect. \ref{sec:discussion}. 

% \par In this work, We calculate the constraint on $H_0$ via the GW emission of the EMRIs with the nuclear transients as EM counterparts. We suppose a relation between EMRI and the EM nuclear transients, including QPEs,QPOs and TDEs, as bright sirens. We calculate the constrain on $H_0$ with plausible bright sirens. The peculiar velocities of the host galaxies of the EMRIs are considered. We found that the fractional error of $H_0$ can be constrained at \textcolor{red}{(1.7\% or 3.7\%)}, depending on QPE models, by QPEs as bright sirens in 4-years observation of LISA. If QPO sources and TDE sources are included to constrain $H_0$, the uncertainty on $H_0$ would be constrained to \textcolor{red}{1.2\% or 1.6\%} in the 2030s.

\section{Methods}\label{sec:Method}
\subsection{Orbital evolution}

%%%%%%%%%%%%%%%%%%%%%%%%%%
% move this part to appendix

%%%%%%%%%%%%%%%%%%%%%%%%%%
\par We model the secular orbital evolution under two distinct channels—(i) a stripping scenario, in which a white dwarf (WD) on an extremely eccentric orbit (eccentricity $e\gtrsim 0.9$) is tidally stripped at each periapsis passage, and (ii) an orbiter–disk collision scenario, in which a compact companion on a circular orbit ($e=0$) intersects a slightly tilted accretion disk twice per orbit. 
\par We assume that the orbital period $P_\text{orb}$ of the system tracks the observed recurrence time of the QPE $T_\text{rec}$, which is theoretically verified in Appendix \ref{app:Trec_Porb}. This assumption leads to $\dot P_\text{orb}\simeq \dot T_\text{rec}$ for the stripping scenario and $\dot P_\text{orb}\simeq 2\dot T_\text{rec}$ for the orbiter-disk collision scenario.
% \par Orbital evolution modeling employs distinct approaches for the stripping and orbiter-disk collision scenarios. For the stripping scenario, it posits a white dwarf (WD) companion on an extremely eccentric orbit (eccentricity $e\gtrsim 0.95$) with a close pericenter around an SMBH with mass $M$ \citep{kingGSN069Tidal2020,zhaoQuasiperiodicEruptionsHelium2022,wangModelPossibleConnection2022}. Whether the recurrence time $T_\text{rec}$ of the QPE matches the orbital period $P_\text{orb}$ is uncertain. There is one eruption per orbit under the stripping scenario. Thus, the relation between $T_\text{rec}$ and $P_\text{orb}$ should be:

% Thus, we assume that the orbital period decay tracks the recurrence time decay.
% % \subsection{Orbital evolution for both models}\label{sec:orb-evo}  
\subsubsection{Orbital decay and evolution for the stripping scenario}\label{sec:orb-evo-stripping}
% 先写orbiter-disk collision model的
% Orbital evolution modeling employs distinct approaches for the stripping and orbiter-disk collision scenarios. For the stripping scenario, it posits a white dwarf (WD) companion on an extremely eccentric orbit (eccentricity $e\gtrsim 0.95$) with a close pericenter around an SMBH with mass $M$ \citep{kingGSN069Tidal2020,zhaoQuasiperiodicEruptionsHelium2022,wangModelPossibleConnection2022}. Whether the recurrence time $T_\text{rec}$ of the QPE matches the orbital period $P_\text{orb}$ is uncertain.

\par The multiple harmonics GW emission in a highly eccentric orbit drives rapid orbital period decay due to orbital shrinking and circularization. In addition, the QPE originates from the periodic mass transfer (MT) from the WD to the central SMBH. Thus, both the MT and the GW affect the orbital evolution.
\par The MT effect can be described by the penetration factor $\beta=R_t/R_p$, where $R_t$ is the disrupted radius, defined by the WD mass $\mu_\text{WD}$ and the WD radius $R_\text{WD}$, namely, $R_t=R_\text{WD}(M/\mu_\text{WD})^{1/3}$; and $r_p$ is the pericenter radius. The $R_\text{WD}$ can be described by the mass-radius relation \citep{zalameaWhiteDwarfsStripped2010}:
\begin{equation}
    R_\text{WD}=9\times10^8\mathrm{cm} ~\left(\frac{M}{M_\odot}\right)^{-1/3}\left(1-\frac{M}{M_\text{ch}}\right)^{0.447}
    \label{eq:rwd}
\end{equation}
where $M_\text{ch}$ is the Chandrasekhar mass.
%%%%%%%%%%%%% 要加上R_WD 的定义

\par The MT occurs if $\beta$ is greater than a value $\beta_0$, representing that the WD fills the Roche lobe. $\beta_0$ is about 0.5 for a non-rotating, cool WD, but rapid spin increases centrifugal force at the surface of the white dwarf, extending $\beta_0$ to $0.3-0.5$ \citep{kruszewskiExchangeMatterClose1963,sepinskyEquipotentialSurfacesLagrangian2007}. A larger $\beta$ indicates stronger MT, and the WD is disrupted when $\beta$ is greater than 1. The MT should affect the orbital evolution of the EMRI. The characteristic timescale of the MT could be represented by:

\begin{equation}
    \left\langle\frac{\dot \beta}{\beta}\right\rangle_{\text{MT}}\simeq\frac{\dot \mu_\text{WD}}{\mu_\text{WD}}\left(\zeta -\frac 1 3 \right) \label{eq:betaMT}
\end{equation}
where $\zeta\sim -1/3$ is the mass-radius index of the WD, giving by $R_\text{WD}\propto M^\zeta$ \citep[see][for details]{wangOrbitalEvolutionTidally2024}. The $\dot \mu_\text{WD}$ is the mass loss rate of the WD, which could be given by \citep{chenTidalStrippingWhite2023}:

\begin{equation}
    {\dot \mu_\text{WD}} \simeq -4.8\frac{\mu}{P_\text{orb}}\left[1-\left(\frac{\mu_\text{WD}}{ M_{\mathrm{ch}}}\right)^{4 / 3}\right]^{3 / 4}\left(1-\frac{\beta_{0}}{\beta}\right)^{5 / 2} \label{eq:mudot}
\end{equation}

\par The orbital evolution of the EMRI is also affected by gravitational radiation.
% In our main analysis, we calculate the orbital evolution by gravitational radiation in 5PN accuracy by \textsc{FEW}. 
Here, we adopt the 0-PN approximation (dominated by quadrupole emission) for simplicity to demonstrate the characteristic timescale of the GW: 
\begin{equation}
    \begin{aligned}
        \left\langle\frac{\dot{\beta}}{\beta}\right\rangle_{\mathrm{GW}}\simeq &\frac{64}{5(1+e)}\frac{G^{3}\mu_\text{WD} M^2}{c^{5}a^{4}\left(1-e^{2}\right)^{5/2}}\\&\left(1-\frac{7}{12}e+\frac{7}{8}e^{2}+\frac{47}{192}e^{3}\right)
    \end{aligned}
     \label{eq:betaGW}
\end{equation}
where $c$ is the speed of light and $a$ is the semi-major axis.
\par The positive sign of both $\langle \dot \beta/\beta\rangle_{\text{MT}}$ and $\langle \dot \beta/\beta\rangle_{\text{GW}}$ indicates the both MT and GW enhance MT. For a typical EMRI system for QPE with SMBH mass $M\sim 10^5M_\odot$, WD mass $\mu_\text{WD}\sim 0.5M_\odot$ for the stripping scenario, the timescale of gravitational radiation is significantly faster than that of MT when $\beta\simeq \beta_0$, with $\langle \dot \beta/\beta\rangle_{\text{MT}}\sim \mathcal O(10^{-17}) \text{s}^{-1}$ and $\langle \dot \beta/\beta\rangle_{\text{GW}} \sim O(10^{-12})\text{s}^{-1}$. This indicates that gravitational radiation dominates orbital evolution at the initial stage. When the $\beta$ evolves to a critical value $\beta_\crit$ at which $\langle \dot \beta/\beta\rangle_{\text{MT}}\simeq\langle \dot \beta/\beta\rangle_{\text{GW}}$, the GW emission balances with the MT, where the orbital period decay stops. After this equilibrium, the MT becomes dominant, and the orbital period increases instead, if the disk friction is not considered (see Sect. \ref{sec:friction}). Thus, the observed evolution of the recurrence time decay is naturally explained as the transition from the GW-dominated phase to the MT-dominated phase.

\par We define the stage from the initial penetration factor $\beta_\text{init}$ to $\beta_\crit$ as the GW-dominated phase, with its duration:
\begin{equation}
    \tau_\text{GW-d}\sim \left\langle \frac{\beta_\text{init}-\beta_\crit}{\dot \beta_{\text{GW}}}\right \rangle
    \label{eq:tau_GW-d}
\end{equation}

Within the GW-dominated phase, the GW emission increases as the orbit evolves. However, the MT dominates after this timescale, increasing the semi-major axis and eccentricity. Thus, we only consider the orbital evolution caused by GW within the GW-dominated phase.

%% 说明我们只考虑GW dominated 期间的，因为超过这个期间，轨道将会越来越大，更不能被探测到。
%%% 说明我们用0PN的estimate GW 和 MT 的时标，结果consist with xian chen's result (考虑MT的演化)， 但是我们用FEW算演化 （高离心率），考虑谐波，更精准，然后考虑谐波的影响在discussion

Considering MT in GW detection is out of the scope of this study. \cite{yeObservingWhiteDwarf2023} demonstrates that the MT decreases the GW amplitude, leading to a more difficult to detect. They show an optimistic detection prospect after the GW-dominated phase, considering MT and tidal deformation of WD. They show detection horizons for detecting a WD with MT in a value of about 200 Mpc with an orbital period less than 2.5 hours. The conclusion may benefit the GW detection for subsequently found QPE sources, but not the current ones, since they are unlikely to evolve to such a short orbital period in the MT-dominated phase.

\subsubsection{Orbital decay for the orbiter-disk collision scenario}\label{sec:orb-evo-collision}
% \par The orbiter-disk collision model posits the inspiral in a near-circular orbit ($e\sim 0$), with the orbiter collapsing the disk twice a orbit.

%% 假设符合理论与观测推断 citep{Eppur si muove} 里面给出了orbital 的拟合
% Given the energy budget requirement that the QPE energy should be far smaller than the system energy loss, 
\par The gravitational dissipation timescale in circular orbit ($e= 0$) is notably slow for explaining the observed recurrence time decay. This suggests that other effects shorten the orbital period $P_\text{orb}$, like the fast orbital evolution due to energy loss in the orbiter-disk interactions \citep{syerStarDiscInteractions1991,macleodEffectStarDisk2020,linialPeriodEvolutionRepeating2024}
\subsection{GW waveform}\label{sec:GW}
\par We follow the derivation in \cite{zhouProbingOrbitsStellar2024a}, which is described as follows. The orbital period decay can be expressed by the energy loss per collision:
\begin{equation}
\dot P_{\text{orb}}=\frac{\mathrm d P_\text{orb}}{\mathrm d E_{\text{orb}}}\dot E_{\text{orb}}=\frac{\mathrm d P_\text{orb}}{\mathrm d E_{\text{orb}}}\frac{\delta E}{2P_\text{orb}}
\label{eq:pdot-E}
\end{equation}
where $\delta E_\text{orb}$ represents the energy loss per collision, and the last term represents the average energy loss in the whole orbit with two collisions. Given Kepler's law $P_{\text{orb}}^2=4\pi^2r^3/(GM)$ and the orbital energy $E_\text{orb}=GM\mu/(2r)$, the differential relation between $E_\text{orb}$ and $P_\text{orb}$ is 
\begin{equation}  
    \frac{\mathrm d P_\text{orb}}{\mathrm d E_{\text{orb}}}=\frac{3\pi^2 G^2M^2\mu^3}{4P_\text{orb} E_{\text{orb}}^4}
    \label{eq:dedp}
\end{equation}

Substitute Eq. \eqref{eq:dedp} to Eq. \eqref{eq:pdot-E}, one can simplify the orbital period decay:
\begin{equation}
    \dot P_\text{orb}=-3\frac{\delta E}{E_\text{orb}}
\end{equation}

% \par The energy loss $\delta E$ depends on the properties of the disk and the nature of the orbiter. 

% We discuss the properties of the disk Sect. \ref{sec:disk}.  

\par The orbital energy lost at each collision is supplied by the drag the orbiter experiences while crossing through the disk and therefore depends on the local disk structure. We assume a geometrically thin $\alpha-$disk around an SMBH with a mass of $M$ in a steady accretion rate $\dot M=\dot m \dot M_\text{Edd}$ where $\dot M_\text{Edd}$ is the Eddington accretion rate, defined by radiation efficiency $\eta$, speed of light $c$ and Eddington luminosity $L_\text{Edd}$ with expression $L_\text{Edd}=\eta \dot M_\text{Edd}c^2$. During the QPE quiescent phase, the bolometric luminosity is typically low, $L_{\rm bol}\sim10^{42}\ {\rm erg,s^{-1}}$ \citep{arcodiaXrayQuasiperiodicEruptions2021,arcodiaTickingAwayLongterm2024}, implying a sub-Eddington flow with $\dot m\simeq L_{\rm bol}/L_{\rm Edd}\sim 0.05$. We therefore adopt a fiducial thin disk with $\dot m=0.05$.

\par We assume that all the orbiter–disk impacts occur in region B of the disk for all sources, based on the orbital period (see Appendix \ref{app:disk_structure} for details).
In region B, where the gas pressure dominates over the radiation pressure, the geometry of the disk can be characterized by the thickness-radius ratio (H/R) \citep{lipunovaStandardModelDisc2018}:
\begin{equation}
\begin{aligned}
    \left(\frac{H}{R}\right)_{B}\simeq &~ 9.3\times 10^{-3}\left(\frac{\alpha}{0.1}\right)^{-1/10} \left(\frac{\dot m}{0.05}\right)^{1/5}\\
    &\left(\frac{M}{10^5 M_\odot}\right)^{-7/20}\left( \frac{R}{100 R_g} \right)^{1/20}
\end{aligned}
\label{eq:hr_zoneB}
\end{equation}

In addition, the surface density $\Sigma_B$ is given in \cite{shakuraBlackHolesBinary1973}:
\begin{equation}
\begin{aligned}
     \Sigma_{B}\simeq &~ 3.0\times10^{5}\mathrm{g/cm}^{2} \left( \frac{\alpha}{0.1} \right)^{-4/5}\left( \frac{\dot m}{0.05} \right)^{3/5}\\
     & \left( \frac{M}{10^5M_\odot} \right)^{1/5}\left( \frac{R}{100 R_g} \right)^{-3/5}
\end{aligned}
    \label{eq:sigma_zoneB}
\end{equation}

%%%%%%%%%%%%%%%%%%%%%%%%%%%%%%%%%%%%%%%
%% move the description about the accretion disk region to the Appendix
%%%%%%%%%%%%%%%%%%%%%%%%%%%%%%%%%%%%%%%

% \par The collision drives orbital decay through drag exerted by the gas. 
Two types of drag forces are usually discussed in the realm of QPE. One is the gravitational drag of the disk \citep{rephaeliFlowMassiveObject1980,ostrikerDynamicalFrictionGaseous1999}:  
\begin{equation}
    F_{\text{grav}}\simeq -4\pi \ln \Lambda \frac{G^2 \Sigma  \mu^2}{Hv_{\text{rel}^2}}
    \label{eq:grav}
\end{equation}
where $H$ is the local scale height of the disk, $v_\text{rel}$ is the relative velocity between the orbiter and the disk, and $\ln \Lambda=\ln(b_{\max}/b_{\min})$ is the Coulomb logarithm, with $b_{\max{}}/b_{\min{}}$ the maximum/minimum cutoff distance associated with the interaction. For simplicity, we fix $\ln \Lambda=10$, as a typical value \cite{zhouProbingOrbitsStellar2024a,xianSecularPeriodicEvolution2025}. Eq. \eqref{eq:grav} follows from the standard gaseous dynamical-friction expression $F_{\rm grav}\propto \rho(G\mu)^{2}/v_{\rm rel}^{2}$ after substituting the gas density expression $\rho\sim\Sigma/H$.

\par Another one is the hydrodynamic friction force \citep{macleodEffectStarDisk2020,linialPeriodEvolutionRepeating2024}:
\begin{equation}
    F_\text{fric}\simeq -C_D\pi R_\text{eff}^2\frac{\Sigma}{H} v_\text{rel}
    \label{eq:friction}
\end{equation}
where $C_D$ is a dimensionless drag coefficient, $R_\text{eff}$ is the effective cross section of the orbiter.

\par There is ongoing debate as to whether the orbiter is a stellar-mass black hole (sBH) ($\mu\sim 100M_\odot$) \citep{franchiniQuasiperiodicEruptionsImpacts2023}, intermediate-mass black hole (iBH) ($\mu\sim 10^3 M_\odot$) \citep{lamBlackHoleaccretionDisk2025} or a main-sequence star ($\mu \sim 1 M_\odot$) \citep{linialCoupledDiskstarEvolution2024,linialUltravioletQuasiperiodicEruptions2024,linialEMRITDEQPE2023}. The last scenario disfavors for GW studies: star–disk interactions are expected to disrupt the star well before the system reaches the LISA band, rendering the GW signal undetectable \citep{linialEMRITDEQPE2023,linialCoupledDiskstarEvolution2024}. Accordingly, we focus on black-hole companions and analyse two fiducial cases: an sBH with $\mu=100 M_\odot$ and an iBH with $\mu=10^3 M_\odot$, though the BH orbiter may face some challenges (see Appendix \ref{app:bh-disk_challenge} for details).

\par Notably, the effective cross section $R_\text{eff}=2G\mu/v_{\rm rel}^2$ is negligible for a BH. For the parameters relevant to our systems, the hydrodynamic friction contribution is subdominant to gravitational drag, and we neglect it below.  

\subsubsection{sBH-disk collision scenario}

The energy loss in every collision can be expressed by $\delta E=F_{\text{grav}}\Delta L$, where $\Delta L=2H/\sin \theta_{\text{od}}$ is the path length crossing the disk for each collision and $\theta_\text{od}$ is the misalignment angle between the orbital plane and the disk. We assume a small tilt with $\theta_\text{od}$ between the orbiter and the disk, leading to the relative velocity $v_\text{rel}\simeq v_K \theta_\text{od}\sim v_K \sin \theta_\text{od}$, where $v_K=(2\pi GM/P_\text{orb})^{1/3}$ is the Kepler velocity. 
Because $F_{\rm grav}\propto v_{\rm rel}^{-2}$, larger misalignments (larger $v_{\rm rel}$) suppress the drag and hence the decay rate, making it harder to match the observed $\dot P_{\rm orb}$ in the BH–disk collision scenario.

\par Combining Eq. \eqref{eq:pdot-E}, \eqref{eq:grav}, and the conditions we gave above, the orbital decay can be expressed by 
\begin{equation}
   \dot P_\text{orb,sBH}=-24\pi \ln \Lambda \frac{G \Sigma \mu R}{Mv_K^2\sin^3\theta_\text{od}} 
    \label{eq:pdot_sbh}
\end{equation}

\par Substituted Eq. \eqref{eq:sigma_zoneB} into Eq. \eqref{eq:pdot_sbh}, one can derive the numerical form of the orbital decay rate:

\begin{equation}
\begin{aligned}
     \dot P_\text{orb,sBH}&\simeq -2.6\times 10^{-5} \text{ s s}^{-1} \left(\frac{\alpha}{0.1}\right)^{-4/5}\left(\frac{P}{5\text{h}}\right)^{14/15}  \\
     &\left(\frac{\ln \Lambda}{10}\right) \left(\frac{\dot m}{0.05}\right)^{3/5} \left(\frac{\mu}{100M_\odot}\right) \\
     &\left(\frac{\sin\theta_\text{od}}{0.1}\right)^{-3}\left(\frac{M}{10^5M_\odot}\right)^{-11/15}
     \label{eq:pdot-analytic_sbh}
\end{aligned}
\end{equation}

\subsubsection{iBH-disk collision scenario}
% Additionally, recent research further constrains the radius of the companion and favors a more compact and massive one \citep{guoTestingStardiskCollision2025,lamBlackHoleaccretionDisk2025}, although the low event rate disfavors a massive orbiter with mass $\sim 10^3 M_\odot$ \citep{linialEMRITDEQPE2023}. 
With a BH orbiter with mass of $10^3 M_\odot$, the strong accretion of the orbiter introduces another drag force, namely Bondi–Hoyle–Lyttleton (BHL) accretion drag \citep{leeDynamicalFrictionGas2014}. The accretion can be expressed by
\begin{equation}
    \dot{\mu}_\text{BHL}\simeq \frac{4\pi(G\mu)^{2}\rho}{(v_{\text{rel}}^{2}+c_s^{2})^{3/2}}
    \label{eq:BHL}
\end{equation}
where $c_s$ is the sound speed of the local gas. The orbiter is supersonic for a thin disk, i.e. $v_\text{rel}>>c_s$ \citep{frankAccretionPowerAstrophysics2002}. The accretion leads to a reactive force, defined by 
\begin{equation}
\begin{aligned}
    F_\text{BHL}&=\dot \mu_\text{BHL} v_\text{rel}\simeq \frac{4\pi(G\mu)^{2}\Sigma}{Hv_{\text{rel}}^{2}}\\
    &= \frac{F_\text{grav}}{\ln \Lambda}
\end{aligned}
    \label{eq:Facc}
\end{equation}
With $\ln \Lambda=10$, the orbital decay is still dominated by the gravitational drag in the iBH-disk collision scenario, with an expression $\dot P_\text{orb, iBH}=1.1\dot P_\text{orb, sBH}$.

\subsubsection{Orbital decay evolution for the BH-disk collision scenario}
\par Observations indicate that the orbital-period decay evolves in several QPE systems \citep{arcodiaTickingAwayLongterm2024,pashamAliveStronglyKicking2024a,pashamAliveBarelyKicking2024}, with direct implications for GW detectability. This evolution may be explained by the Lense-Thirring (LT) precession of the orbiter and the disk \citep{arcodiaTickingAwayLongterm2024,zhouProbingOrbitsStellar2024a}. Notably, because warps in a viscous disk diffuse under the vertical viscosity, the disk generally precesses more slowly than the orbiter, leading to differential precession, and may ultimately align with the SMBH equatorial plane in (see Sect. \ref{sec:aligment}). Denoted the LT precession frequency of the orbiter as $\Omega_\text{LT,orb}$, the LT precession frequency $\Omega_\text{LT,disk}$ for the disk can be expressed by, $\Omega_\text{LT,disk}= \kappa\Omega_\text{LT, orb}$, where $\kappa$ is a proportionality constant, adopting from \cite{arcodiaTickingAwayLongterm2024}. The differential precession changes the misalignment angle $\theta_{\rm od}(t)$ between the orbital plane and the disk, which in turn modulates $\dot P_{\rm orb}$ via Eq. \eqref{eq:pdot-analytic_sbh}.

\par The local LT precession frequency in the linear regime can be expressed by \citep{franchiniLENSETHIRRINGEFFECTACCRETING2017}    
\begin{equation}
    \Omega_{\mathrm{LT}}=\frac{2ac^3}{GMr^3}
    \label{eq:LT_f}
\end{equation}
where $r=R/R_g$ and $a$ is the dimensionless spin of the SMBH. Thus, the LT timescale should be 
\begin{equation}
\begin{aligned}
    \tau_\text{LT}&=\frac{2\pi}{\Omega_\text{LT}}\\
    &\simeq 3.32 \text{ yr } \left(\frac{M}{10^5M_\odot}\right)^{-1}\left(\frac{P}{5\text{h}}\right)^2\left(\frac{0.5}{a}\right)
\end{aligned}
    \label{eq:tauLT}
\end{equation}
In this scenario, the frequency of the differential precession is $\Delta\Omega \equiv \Omega_{\rm LT,orb}-\Omega_{\rm LT,disk}=(1-\kappa)\Omega_{\rm LT,orb}$. Thus, the characteristic period over which $\theta_{\rm od}$—and hence $\dot P_{\rm orb}$—is  $\tau_p=\tau_\text{LT}/(1-\kappa)$. For $\kappa$ moderately larger than unity, $\tau_{p}$ is of order $\tau_{\rm LT}$; in our fiducial examples we therefore set $\tau_{p}\approx\tau_{\rm LT}$ for simplicity.

\subsection{GW waveform}
%%% 差了个参数怎么取，比如距离，参数依赖等等

\par In the two scenarios, QPEs involve two types of inspiral systems, namely EMRI and IMRI, depending on the mass ratio $q\equiv \mu/M$. For the stripping scenario, the QPEs involve a WD orbiter with the mass ratio $q \lesssim 10^{-4}$, leading to EMRI systems. The eccentric EMRI introduces high-frequency harmonics beyond the quadrupole emission. Thus, we utilize \texttt{FastEmriWaveform} (\texttt{FEW}) framework to compute orbital evolution and gravitational wave emission  \citep{katzFastEMRIWaveformsNewTools2021}. This framework can simulate orbital evolution through GW emission and rapidly generate accurate EMRI waveforms with 5-PN maximum accuracy, as described in an improved augmented analytic kludge model. In our implementation, seven parameters are used: $\{M,\mu,a,p_0,e_0,\iota,d_L\}$, where $M$ and $\mu$ are the SMBH and orbiter masses, $a$ is the dimensionless, $p_0$ and $e_0$ are the initial semi-latus rectum and eccentricity, $iota$ is the orbital inclination, and $d_L$ is the luminosity distance. The definitions of these parameters are listed in Tab. \ref{tab:parameter_notation}.

\par In the orbiter-disk collision scenario, the orbiter is a sBH ($\mu=100M_\odot$) or iBH ($\mu=10^3M_\odot$). Both orbiters indicate an IMRI system with mass ratio $q \gtrsim 10^{-4}$. For these systems, we model the signal with \texttt{IMRPhenomD} \citep{khanFrequencydomainGravitationalWaves2016,husaFrequencydomainGravitationalWaves2016}, which is designed for non-precessing systems on circular orbit. This model provides the waveform in (2,2) harmonic mode, which dominates in such a system, in the expression:
\begin{equation}
\begin{aligned}
\tilde h_+(f)&=\frac{1+\cos^2 \iota}{2} ~\mathcal A_{2,2}(f) \exp\big[\mathrm i \Psi_{2,2}(f)\big]\\
\tilde h_\times(f) &= -\mathrm i \cos \iota ~\mathcal A_{2,2}(f) \exp\big[\mathrm i \Psi_{2,2}(f)\big]
\end{aligned}
\label{eq:imr_waveform}
\end{equation}
where $\mathcal{A}_{2,2}$ and ${\Psi}_{2,2}$ are the amplitude and the phase in (2,2) harmonic mode. The frequency-dependent waveform can be calculated by $\tilde h(f) =\tilde h_+(f)-\text i \tilde h_\times(f)$.  This waveform model relies on fewer parameters, i.e., $\{M,\mu,\iota,d_L\}$. 

\begin{table}[htbp]
\caption{The notations and their definitions of parameters in waveform models} \label{tab:parameter_notation}

\hspace{-1cm}
\begin{tabular}{@{}cl@{\hspace{1.5cm}}l@{}}
\toprule
notation &  & definition \\ \midrule
$M$ &  & The mass of the central SMBH \\
$\mu$ &  & The mass of the orbiter \\
$d_L$ &  & The luminosity distance \\
$\iota$ &  & The orbital inclination \\
$a$ &  & The dimensionless SMBH spin \\
$p_0$ &  & The initial semi-latus rectum of the orbit\\
$e_0$ &  & The initial eccentricity of the orbit\\ \bottomrule
\end{tabular}%
\end{table}

\subsection{Parameter estimation: Fisher matrix}\label{sec:parameter}
%% 这里的参数估计用传统的并合，不要用现在写的这种
\par If a QPE source produces a detectable GW signal in the 2030s after secular evolution, it would constitute a `bright siren', enabling a direct constraint on the $H_0$. In this framework, the redshift is obtained from the EM counterpart by identifying the host galaxy, while the distance is inferred from the GW signal. Since QPEs can uniquely identify their host galaxies, the sky location of the host galaxies can be measured with negligible error. Thus, we assume that the sky locations are fixed throughout our analysis. 

\par The distance is estimated with matched filtering. For two Fourier-domain waveforms $\tilde h(f)$ and $\tilde g(f)$, the match-filter is described by the inner product between the signals, which is defined by
\begin{equation}
    (\tilde h|\tilde g)=2\Re \left[\int \frac{\tilde h(f) \tilde g(f)^*+\tilde h(f)^*\tilde g(f)}{S_n(f)}\mathrm d f\right],
    \label{eq:inner_product}
\end{equation}
where $S_n(f)$ is the one-sided noise spectrum \citep{babakLISASensitivitySNR2021}. And the signal-to-noise ratio (SNR) of a GW signal is characterized by the square root of the inner product of itself, namely, $\text{SNR}= \sqrt{(\tilde h|\tilde h)}$. 

\par We adopt the Fisher information matrix (FIM) to calculate Parameter uncertainties. As it is described in Sect. \ref{sec:GW}, the waveform model depends on seven parameters $\Xi\in\{M,\mu,a,p_0,e_0,\iota,d_L\}$ (adopted for \texttt{FEW}) or four parameters $\Xi\in\{M,\mu,\iota,d_L\}$ (adopted for \texttt{IMRphemonD}). The FIM is 
\begin{equation}
    \Gamma_{ij}=\left( \frac{\partial \tilde h}{\partial \Xi_i} \middle | \frac{\partial \tilde h}{\partial \Xi_j}\right)
\end{equation}
where subscripts i, j represent \textit{i}-th and \textit{j}-th parameters.
\par The covariance of the parameters can be represented by the FIM, i.e., $\text{Cov}_{ij}(\Xi)=1/\Gamma_{ij}(\Xi)$. So the $1\sigma$ uncertainty on the parameter $\Xi_i$ is $\Delta { \Xi_i}=\sqrt{\text{Cov}_{ii}(\Xi)}=1/\sqrt{\Gamma_{ii}(\Xi)}$. 

\par Propagating uncertainties in the low-redshift Hubble law (Eq.~\ref{eq:hubble}), and and neglecting any covariance between $v_r$ and $d_L$, the fractional uncertainty on $H_0$ is 
\begin{equation}
    E(H_0)=\frac{\Delta H_0}{H_0}=\sqrt{\left(\frac{\Delta v_r}{v_r}\right)^2+\left(\frac{\Delta d_L}{d_L}\right)^2}
    \label{eq:uncertainty_h0}
\end{equation}
where the recessional-velocity uncertainty $\Delta v_r$ is dominated by the line-of-sight mean peculiar velocity $\langle{v}\rangle$. This velocity is set as a typical value of $\langle{v}\rangle =300 \text{km/s}$ \citep{adePlanckIntermediateResults2014,carrickCosmologicalParametersComparison2015,grazianiPeculiarVelocityField2019}.

%%%%%%% 把结果里面QPO都去掉，放到discussion 里面，在result里面就讨论regular和uncertain的
\section{Result}\label{sec:result}
\par We evaluate GW detectability under two representative QPE frameworks: a stripping scenario and an orbiter–disk collision scenario, each incorporating secular orbital evolution. Notably, not all reported QPEs are suitable for these prescriptions, and our selection reflects the quality and behavior of the timing data.
\par Observationally, two QPEs—GSN 069 and eRO-QPE2—have exhibited approximately secular decreases in recurrence time. For eRO-QPE2, the decay trend appears to flatten and may effectively vanish on a $\sim 3$-yr baseline \citep{arcodiaTickingAwayLongterm2024}, while no statistically significant secular evolution is currently established for GSN 069 \citep{miniuttiEppurSiMuove2025}. In what follows, we attribute the recurrence time decay to the orbital decay, and we interpret changes in the decay rate as the MT effect in the stripping scenario and LT precession in the orbiter-disk collision scenario.
\par In contrast, eRO-QPE1 and RX J1301.9+2747 show irregular or non-monotonic timing behavior, without a stable decay evolution \citep{pashamAliveBarelyKicking2024,giustiniFragmentsHarmonyApparent2024}. The origin of the irregular timing pattern is model-dependent and presently ambiguous. Thus, we do not assess their GW detectability in this paper.

\par Several additional sources (eRO-QPE3, eRO-QPE4, AT2019qiz, AT2022upj, and ZTF19acnskyy) lack secular monitoring to establish secular trends in the orbital period. Among them, eRO-QPE3 and eRO-QPE4 offer the only plausible near-term prospect for entering LISA's band due to its comparatively short period. The other sources would not reach the sensitive band of LISA even after yearly orbital evolution in a non-physical rate ($\sim 10^{-4}\text{ s s}^{-1}$). Therefore, we will discuss the GW detectability of eRO-QPE3 and eRO-QPE4 for reference.

\par The Europe-led GW detection mission, LISA, is planned for launch in 2035 and is designed to operate for 4.5 years, with consumables allocated to support a potential extension to ten years \citep{colpiLISADefinitionStudy2024}. Thus, we calculate a 15-year orbital evolution for each source we considered in two scenarios. In addition, we utilize \texttt{FEW} toolkit to calculate the GW emission for the EMRI system in the stripping scenario and \texttt{IMRphenomD} for the IMRI system in the orbiter-collision scenario. 
%%% 注意， 讲模型的时候要将EMRI和IMRI， 以及讲用什么来计算

\subsection{The detectability of the stripping scenario}
\par
\begin{figure*}[htbp]
    \centering
    \includegraphics[width=\linewidth]{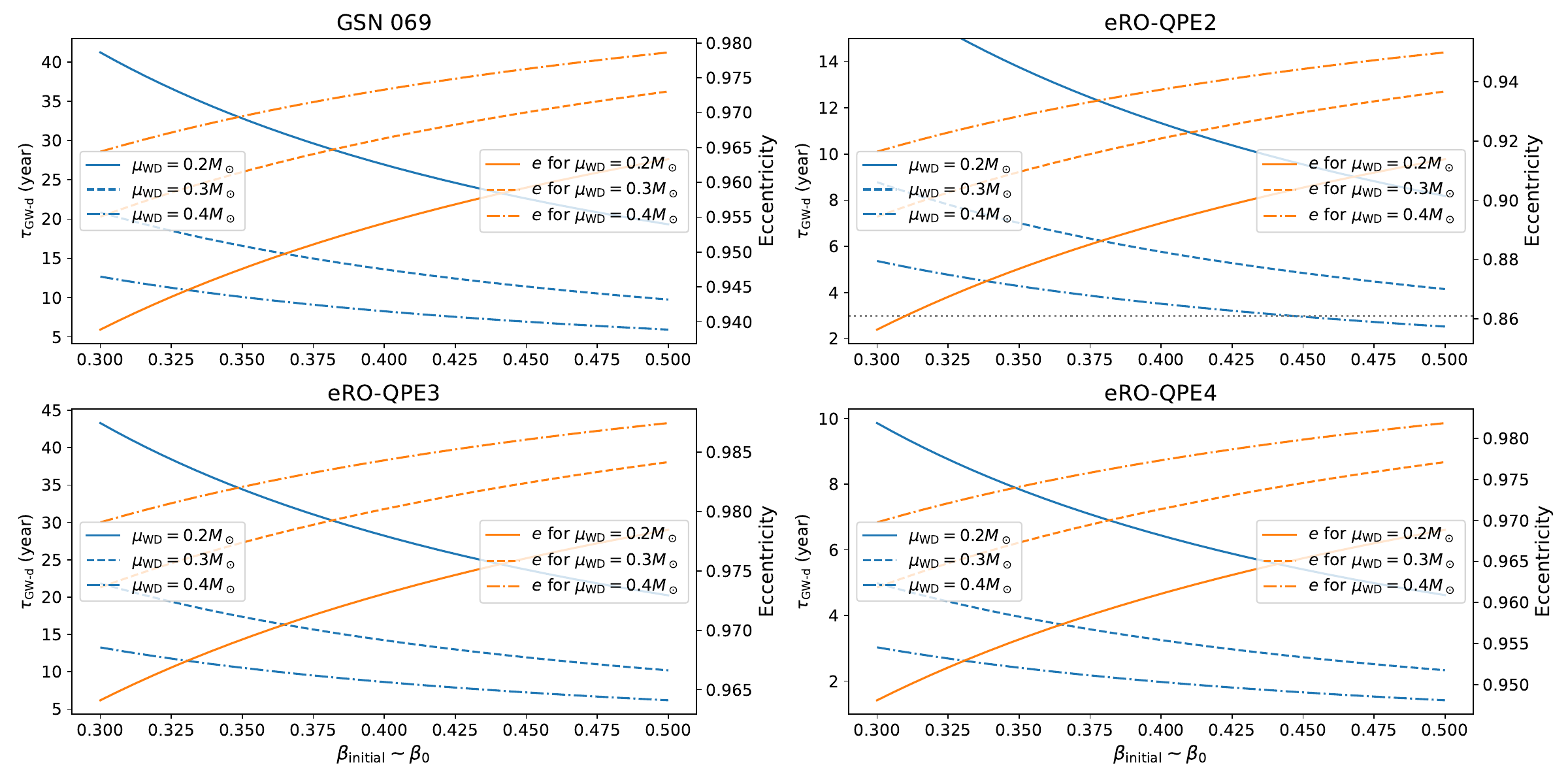}
    \caption{The dependence of orbital eccentricity $e$ and $\tau_\text{GW-d}$ with respect to $\mu_\text{WD}$ and $\beta_\text{init}$. With a larger $\beta_\text{init}$, the $\tau_\text{GW-d}$ increases and the $e$ decreases, while the situation is opposite with a larger $\mu_\text{WD}$. The gray dashed line in the top-right figure represents the $\tau_\text{GW-d}=3\text{ yr}$ of eRO-QPE2, indicated by the observation.}
    \label{fig:timescale_beta}
\end{figure*}

 In the stripping scenario, we assume that the MT is neglected during the early, GW-dominated phase, where the GW emission drives the orbital decay. 
% In this interval, the orbital evolution is dominated by GW emission, yielding $\dot P_\text{orb}<0$. Once MT is set in, it reverses the sign of the orbital decay rate and drives orbital expansion, i.e. $\dot P_\text{orb} >0$, at which point the GW signal rapidly weakens and becomes effectively undetectable. Observationally, the decay of the recurrence time then vanishes (i.e., approaches zero), as seen in eRO-QPE2. 
The duration of the GW-dominated phase $\tau_\text{GW-d}$ is determined by the initial $\beta$ and the mass of the WD $\mu_\text{WD}$. We calculate the $\tau_\text{GW-d}$ across $\beta_\text{init}$ and $\mu_\text{WD}$, as demonstrated in Fig. \ref{fig:timescale_beta}, using the fiducial parameters described in Tab. \ref{tab:stripping_parameter}. For simplicity, we set the spin of the central SMBH to 0.5 and the inclination of the orbit to 0. The orbital eccentricity $e$ can be calculated by $P_\text{orb}$, $\mu_\text{WD}$ and $\beta$, and is reported alongside $\tau_{\rm GW\mbox{-}d}$ in Fig. \ref{fig:timescale_beta}. Increasing $\beta_\text{init}$ lengthens $\tau_\text{GW-d}$ and yields lower $e$, whereas increasing $\mu_\text{WD}$ has the opposite effect.

\par For eRO-QPE2, the observations indicate $\tau_\text{GW-d}\sim 3 \text{ yr}$ since the measured $\dot P_\text{orb}$ approaches zero in about 3-year timescale \citep{arcodiaTickingAwayLongterm2024}. Fig. \ref{fig:timescale_beta} indicates the system with initial parameters $\mu_\text{WD}=0.4M_\odot, \beta_0=0.45$ can match the observation. The initial orbital decay rate $\dot P_\text{orb}\sim \mathcal O(-10^{-5}) \text{ s s}^{-1}$ verifies the explanation. In this scenario, the source will already be MT-dominated during the LISA epoch, and its GW emission is expected to be too weak for detection.

\begin{table}[h]
\centering
\caption{The parameters for the QPEs in the stripping scenario} \label{tab:stripping_parameter}
\begin{tabular}{@{}llcccc@{}}
\toprule
 &  & $M~(M_\odot)$ & $P_\text{orb} ~(\text{h})$ & $\mu_\text{WD}~(M_\odot)$ & $\beta_\text{init}$ \\ \midrule
GSN 069 &  & $3\times 10^5$ & 9 & 0.2-0.4 & 0.3-0.5 \\
eRO-QPE2 &  & $1\times 10^5$ & 2.5 & 0.2-0.4 & 0.3-0.5 \\
eRO-QPE3 &  & $2\times 10^6$ & 20 & 0.2-0.4 & 0.3-0.5 \\
eRO-QPE4 &  & $2\times 10^7$ & 11.5 & 0.2-0.4 & 0.3-0.5 \\ \bottomrule
\end{tabular}%
\end{table}

\par
\begin{figure}[h]
    \centering
    \includegraphics[width=\linewidth]{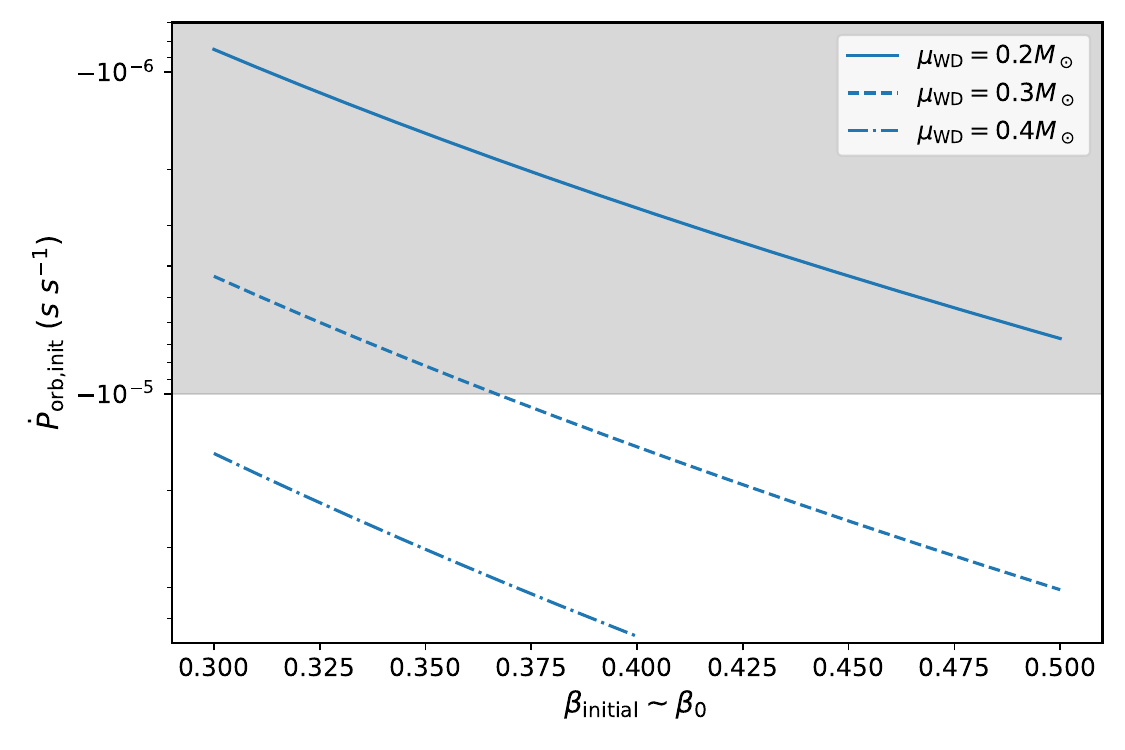}
    \caption{The initial orbital decay $\dot P_\text{orb,init}$ with respect to $\mu_\text{WD}$ and $\beta_\text{init}$. With a larger $\beta_\text{init}$ and a larger $\mu_\text{WD}$, the $|\dot P_\text{orb,init}|$ increases. The observation indicates that $\dot P_\text{orb,init} \sim -10^{-5}$  for GSN 069. The gray block represents the parameter space that failed to match the observation. The constraint on $\dot P_\text{orb,init}$ leads high $\beta_\text{inital}$ and low $\mu_\text{WD}$.}  
    \label{fig:timescale_beta_gsn069}
\end{figure}

 For GSN 069, Fig. \ref{fig:timescale_beta} shows that $\tau_\text{GW-d}$ can reach over 20 years for a sufficiently small $\beta_\text{init}$ and low $\mu_\text{WD}$. However, GSN 069 is further constrained by the observed orbital decay $\dot P_\text{orb}$ with the value $\sim 10^{-5} \mathrm {s~s^{-1}}$ \citep{miniuttiEppurSiMuove2025}. We estimate the initial orbital decay $\dot P_\text{orb,init}$ across $\mu_\text{WD}$ and $\beta_\text{init}$ by \texttt{FEW}. We remind the reader here that the 0-PN approximation (Eq. \ref{eq:betaGW}) is only utilized to estimate $\tau_\text{GW-d}$; the orbital evolution and $\dot{P}{\rm orb,init}$ are obtained from the multi-harmonic EMRI waveform model \texttt{FEW}. 
% More discussion on introducing multi-harmonic emission is described in Sect. \ref{sec:environment}. 
As shown in \ref{fig:timescale_beta_gsn069}, the $|\dot P_\text{orb,init}|$ increases with larger $\beta_\text{init}$ and $\mu_\text{WD}$. Matching $\dot P_\text{orb,init}\sim -10^{-5}\mathrm{s~s^{-1}}$ favors relatively large $\beta_\text{init}$ ($\gtrsim 0.375$) and low $\mu_\text{WD}$ ($\lesssim 0.3M_\odot$), which in turn imply $\tau_\text{GW-d}<15\text{ yr}$. Under these conditions, GSN~069 is expected to enter the MT-dominated phase in the 2030s, further reducing its GW emission. Even in an optimistic case with $\tau_\text{GW-d}\sim 15\text{ yr}$ ($\beta_\text{init}=0.375, \mu_\text{WD}=0.3M_\odot$) and ignoring MT while computing the waveform, the predicted LISA SNR over a 4-yr observation remains $\ll 1$, resulting in an impossible detection.

\par For eRO-QPE3 and eRO-QPE4, there is no direct constraint on $\dot P_\text{orb}$ yet. However, a necessary (though not sufficient) condition for a LISA detection during the GW-dominated epoch is that the source remains in this phase for $\tau_\text{GW-d}\gtrsim 20\text{ yr}$ (allowing 15-year orbital evolution and 4-year observation). This constraint rules out the GW detectability of eRO-QPE4 according to Fig. \ref{fig:timescale_beta}, which implies that no parameter space is allowed for eRO-QPE4. In contrast, the constraint allows initial orbital parameters that $\beta_\text{init}=0.32$ and $\mu_\text{WD}=0.3M_\odot$ for eRO-QPE3; however, the GW SNR is $\sim 0.06 \ll 1$ in a 4-year observation. 
\par We caution that \texttt{FEW} is not accurate for extreme eccentric orbits, where this framework neglects partial high-order harmonic modes. This caveat does not affect our conclusions here, since stripping-scenario sources are far below detectability in all explored configurations. Moreover, we adopt 0-PN accuracy to estimate $\tau_\text{GW-d}$ for simplicity. \citet{yangImpactMassTransfer2025} indicates an even shorter $\tau_\text{GW-d}$, who consider a higher PN method and WD surface structures. This finding further prevents GW detections for the stripping scenario.   
\par In summary, none of the known QPEs are expected to be detectable by LISA in the 2030s within the stripping scenario. Additional gas dynamical effects, e.g., the friction of the accretion flow that hinders accretion, may extend the $\tau_\text{GW-d}$ with the same $(\beta_\text{init},\mu_\text{WD})$ and thereby improve GW detectability. We explore this possibility in Sect. \ref{sec:friction}.    

\subsection{The detectability of the orbiter-disk scenario}
In this scenario, QPEs are powered by repeated impacts between the orbiter and the accretion disk, which extract orbital energy and angular momentum and therefore drive a secular decrease in the orbital period, $\dot P_\text{orb}< 0$. Two orbiters are considered, (i) a sBH with mass $\mu=100 M_\odot$, (ii) an iBH with mass $\mu=10^3 M_\odot$. Unless otherwise noted, we fix the following fiducial parameters for simplicity: the dimensionless spin of the central SMBH $a=0.5$, the viscosity factor $\alpha=0.1$, and the dimensionless accretion rate $\dot m = 0.05$.

\subsubsection{The sBH-disk scenario}
\par Using Eq. \eqref{eq:pdot_sbh}, together with the source properties listed in Tab. \ref{tab:sbh-disk_parameter}, we compute the initial orbital decay for each system. We assume that the evolution of the orbital decay originates from the precession of both the orbit and the disk, which modulates the misalignment angle $\theta_\text{od}$. The inclination of the disk $i_\text{d}$ is fixed to $\sim 0.1 \text{ rad}$ and the initial $\theta_\text{od}=0.1\text{ rad}$, corresponding to the maximum orbital decay $\dot P_\text{orb,max}$. As the system precesses, the $\theta_\text{od}$ increases approximately linearly and reaches $\theta_\text{od,max}\simeq 0.3\text{ rad}\simeq 20^\circ$ after half a precession cycle, $\tau_{\rm LT}/2$, at which point the magnitude of the decay reaches its minimum, $\dot P_{\rm orb,min}$. For secular evolution, approximate the time-varying $\dot P_{\rm orb}$ by its cycle-averaged value, $\langle \dot P_{\rm orb}\rangle$. The parameters we used and the calculated orbital decay are summarized in Tab. \ref{tab:sbh-disk_parameter}. The orbital decay of GSN 069 and eRO-QPE2 is consistent with the measured period evolution \citep{arcodiaTickingAwayLongterm2024,pashamAliveStronglyKicking2024,miniuttiEppurSiMuove2025} and theoretical predictions \citep{zhouProbingOrbitsStellar2024}. In particular, for eRO-QPE2, the $\dot P_\text{orb,min}\sim -10^{-7} \text{ s s}^{-1}$ is sufficient to explain the apparent disappearance of the recurrence-time decay. 

\par Assuming a 15-year orbital evolution with constant $\langle \dot P_{\rm orb}\rangle$, the finial orbital period is simply calculated $P_\text{orb,fi}=P_\text{orb}+15 \text{ yr} \times \langle \dot P_\text{orb}\rangle$. For eRO-QPE3 and eRO-QPE4, the mass ratios of both systems ($q \lesssim 10^{-4}$) lie outside the calibration range of \texttt{IMRPhenomD}; we therefore use the \texttt{FEW} framework instead to calculate their GW emission. For GSN 069 and eRO-QPE3, the orbital decay is too slow to move the systems into the LISA sensitivity band over this interval: we find $P_{\rm orb,fin}\simeq 17$~h and $\simeq 39$~h, respectively, corresponding to SNRs $\ll 1$ for a 4-yr observation. Moreover, for eRO-QPE4, the orbital decay is even slower, yielding $P_{\rm orb,fin}\simeq 22.9$~h. Its SNR of this source is 0.5, slightly stronger than that of GSN 069 and eRO-QPE3 due to the harmonic emission, but still falls well below a conservative detection threshold of SNR $\ge 8$. By contrast, the initially compact orbit of eRO-QPE2 yields $P_{\rm orb,fin}\simeq 4.3$ h and an SNR of $\sim 8.5$ for a 4-yr LISA observation, exceeding the threshold and implying a confident gravitational-wave detection. 
\par We note that the ongoing orbiter-disk interaction during the LISA observing window can imprint modulations on the GW phase and amplitude, thereby altering the SNR. Phase drifts and amplitude fluctuations during collisions produce additional high-frequency harmonic modes, which may raise the SNR against simplified templates in the vacuum. We omit these couplings in the SNR calculation for simplicity, but discuss them qualitatively (see Sect. \ref{sec:environment} for details). 

\begin{table*}[htbp]
\centering
\caption{The parameters of QPEs in the sBH-disk scenario}\label{tab:sbh-disk_parameter}

\hspace{-2cm}
% \resizebox{1.2\linewidth}{!}{%
\begin{tabular}{llcccccc}
\hline
 $\mu=100M_\odot$ &  & $M~(M_\odot)$ & $P_\text{orb} ~(h)$ & $\tau_{\rm LT}$ (yr) & $\dot P_{\rm orb, max} ~(\text{s s}^{-1})$ & $\dot P_{\rm orb, min} ~(\text{s s}^{-1})$& $\langle \dot P_{\rm orb}\rangle ~(\text{s s}^{-1})$ \\ \hline
GSN 069 &  & $3\times 10^5$ & 18 & 14.3 & $-3.8\times 10^{-5} $ & $-1.3\times 10^{-6} $ & $-8.9\times 10^{-6}$ \\
eRO-QPE2 &  & $1\times 10^5$ & 5 & 3.3 & $-2.6\times 10^{-5}$ & $-9.1\times 10^{-7} $& $-6.0\times 10^{-6}$ \\
eRO-QPE3 &  & $2\times 10^6$ & 40 & 10.6 & $-1.7\times 10^{-5}$ & $-7.1\times 10^{-7} $& $-4.7\times 10^{-6}$ \\ 
eRO-QPE4 &  & $2\times 10^7$ & 23 & 0.4 & $-2.2\times 10^{-6}$ & $-7.8\times 10^{-8} $& $-5.2\times 10^{-7}$ \\ 
\hline
\end{tabular}%
% }
\end{table*}

\subsubsection{The iBH-disk scenario}
The calculation follows the same formalism as in the sBH–disk case; only the fiducial parameters are changed. Here, the orbiter is taken to be an iBH with a mass of $10^3M_\odot$. Unless stated otherwise, the remaining parameters are identical to those adopted above (e.g., $a=0.5$, $\alpha=0.1$, $\dot{m}=0.05$). From Eq.~\eqref{eq:pdot-analytic_sbh}, the instantaneous decay scales as $|\dot P_{\rm orb}|\propto \mu(\sin\theta_{\rm od})^{-3}$, so increasing the companion mass by a factor of ten would, at fixed geometry, boost the magnitude of the decay by the same factor. To reproduce the observed $\dot P_{\rm orb}$, the geometric parameters must therefore be adjusted.

\par The disk inclination sets the maximum amplitude of the relative misalignment angle $\theta_\text{od}$ driven by coupled LT precession: as the disk inclination increases, the maximum $\theta_\text{od}$ grows, reaching $\theta_{\rm od,max}\sim 90^\circ$ for the disk inclination approach $45^\circ$. Because of the $(\sin\theta_{\rm od})^{-3}$ dependence, larger $\theta_{\rm od}$ reduces both the cycle-averaged decay rate $\langle\dot P_{\rm orb}\rangle$. To reproduce the observed period evolution, we therefore adopt a larger initial misalignment angle and a moderate disk inclination. Specifically, we set $\theta_{\rm od,ini}=0.2\text{ rad}$ and $i_{\rm d}\simeq 0.25 \text{ rad} ~(15^\circ)$. With these choices, $\theta_{\rm od}$ grows during precession to a larger $\theta_{\rm od,max}$, yielding a smaller $\langle\dot P_{\rm orb}\rangle$ that remains compatible with the observations even for $\mu=10^3M_\odot$, as demonstrated in Tab. \ref{tab:ibh-disk_parameter}.
\par Increasing the orbiter mass naturally enhances the GW mission, improving detectability for all sources at fixed period and distance. Nevertheless, after a 15-year orbital evolution, eRO-QPE3 still yields a final orbital period of $P_\text{orb,fi}\simeq 39~\text{h}$, which is too large for a LISA detection. For GSN 069, the higher mass raises the 4-yr LISA SNR to ${\sim}1.5$, though still well below the SNR threshold. By contrast, the SNRs for eRO-QPE2 and eRO-QPE4 increase to ${\sim}28.8$ and ${\sim} 17.7$, respectively (for the latter we compute the signal with \texttt{FEW} owing to the small mass ratio), implying robust detections.

\begin{table*}[htbp]
\centering
\caption{The parameters of QPEs in the iBH-disk scenario}\label{tab:ibh-disk_parameter}

\hspace{-2cm}
% \resizebox{1.2\linewidth}{!}{%
\begin{tabular}{llcccccc}
\hline
$\mu=10^{3}M_\odot$ &  & $M~(M_\odot)$ & $P_\text{orb} ~(h)$ & $\tau_{\rm LT}$ (yr) & $\dot P_{\rm orb, max} ~(\text{s s}^{-1})$ & $\dot P_{\rm orb, min} ~(\text{s s}^{-1})$& $\langle \dot P_{\rm orb}\rangle ~(\text{s s}^{-1})$ \\ \hline
GSN 069 &  & $3\times 10^5$ & 18 & 14.3 & $-4.9\times 10^{-5} $ & $-1.3\times 10^{-6} $ & $-9.9\times 10^{-6}$ \\
eRO-QPE2 &  & $1\times 10^5$ & 5 & 3.3 & $-3.6\times 10^{-5}$ & $-9.8\times 10^{-7} $& $-7.3\times 10^{-6}$ \\
eRO-QPE3 &  & $2\times 10^6$ & 40 & 3.5 & $-2.8\times 10^{-5}$& $-7.6\times 10^{-7} $& $-5.7\times 10^{-6}$ \\ 
eRO-QPE4 &  & $2\times 10^7$ & 23 & 3.5 & $-3.1\times 10^{-6}$& $-8.3\times 10^{-7} $& $-6.3\times 10^{-7}$ \\ 
\hline
\end{tabular}%
% }
\end{table*}

\subsection{Constraint on $H_0$ using QPE-sirens}\label{sec:h0}

\par Because QPEs are plausibly associated with EMRIs/IMRIs that will be observable by LISA, they provide promising bright sirens for measuring the $H_0$. In this framework, the host galaxy---and hence the redshift---is secured by the QPE identification, while the GW encodes the luminosity distance. We therefore estimate $H_0$ constraints for the two QPEs with forecast GW detectability, eRO-QPE2 and eRO-QPE4, propagating both the GW distance error and the redshift uncertainty induced by peculiar velocities.
\par For the sBH-disk collision scenario, eRO-QPE2 yields an $\rm SNR=8.5$ with 4-year LISA observation, leading to a constraint on $H_0$ in fractional uncertainty of 14.9\%. In the iBH-disk collision scenario, the higher companion mass boosts the SNR of eRO-QPE2 to $28.8$ and tightens the constraint to $10.4\%$. In the same iBH-disk scenario, eRO-QPE4 is detectable with an SNR of 21.8 and derives the tightest constraint on $H_0$ of $6.7\%$, which is comparable to other probes \citep{Yu2018,Wu2022,Gao2025}. The tightest constraint is achieved because the larger distance of eRO-QPE4 suppresses the fractional impact of peculiar velocities, despite its lower SNR relative to the eRO-QPE2, thereby improving the $H_0$ inference. A summary is provided in Table~\ref{tab:snr-h0}.
\begin{table}[htpb]
\centering
\caption{The SNR of the bright sirens and their constraint on $H_0$}\label{tab:snr-h0}
\begin{tabular}{@{}lccccc@{}}
\toprule
 & \multicolumn{2}{c}{sBH-disk} &  & \multicolumn{2}{c}{iBH-disk} \\ \cmidrule(lr){2-3} \cmidrule(l){5-6} 
 & SNR & $E(H_0) (\%)$ &  & SNR & $E(H_0) (\%)$ \\ \midrule
eRO-QPE2 & 8.5 & 14.9 &  & 28.8 & 10.4 \\
eRO-QPE4 & $<1$ & - &  & 21.8 & 6.7 \\ \bottomrule
\end{tabular}%
\end{table}

\section{Discussion}\label{sec:discussion}

\subsection{Additional potential effects for GW detectability}
Our main analysis considers simplified models for both scenarios, neglecting complicated effects. These potential effects may affect the orbital evolution and, therefore, the GW detectability. Numerical calculations of these effects significantly complicate the analysis, which is out of the scope of this paper. Thus, we briefly discuss the potential effects here.

\subsubsection{For the stripping scenario: increase $\tau_\text{GW-d}$ by disk friction} \label{sec:friction}
\par In the stripping scenario, the WD on a highly eccentric orbit loses mass at each periapsis passage. The stripped matters circularize and form an accretion disk around the SMBH. In addition, in several systems, the QPEs appear temporally associated with TDE activity\citep{miniuttiAliveKickingNew2023,miniuttiRepeatingTidalDisruptions2023,nichollQuasiperiodicXrayEruptions2024,chakrabortyDiscoveryQuasiperiodicEruptions2025}, indicating that a pre-existed accretion disk is present during the QPE phase. The gaseous disk provides a drag force on the WD, modifies the duration of the GW-dominated phase $\tau_{\text{GW-d}}$, in close analogy with the orbiter–disk collision scenario. Two drag forces are considered, namely $F_\text{grav}$ (Eq. \eqref{eq:grav}) and $F_\text{fric}$ (Eq. \eqref{eq:friction}).
% Two drag forces are considered, namely, the hydrodynamic force or the dynamical friction. 
% The hydrodynamic timescale is :
% \begin{equation}
%     F_\text{HF}\simeq C_D \rho_g v^2_\text{rel} \pi R^2
%     \label{eq:drag_hf}
% \end{equation}
% where $C_D$ is related to the shape of the star and is equal to 1 for a sphere, while $\rho_g$ is the density of the local gas, and $v_\text{rel}$ is the relative velocity of the star with respect to the local gas \citep{szolgyenEccentricityEvolutionGaseous2022}. The dynamical friction is \citep{ostrikerDynamicalFrictionGaseous1999}
% \begin{equation}
%     F_\text{DF} \simeq 4 \pi \rho_g (G\mu_\text{WD})^2 \frac{\mathcal{I}(\mathcal M)}{v^2_\text{rel}}
%     \label{eq:drag_df}
% \end{equation}
% % where $\mathcal M$ is the Mach number, and $\mathcal I$ is the correction factor \citep{ostrikerDynamicalFrictionGaseous1999}.
For convenience, we define a local drag timescale by
\begin{equation}
    \tau_F=\frac{\mu_\text{WD}v_\text{rel}}{F}
    \label{eq:drag_timescale}
\end{equation}

% \par The orbital evolution due to the drag force can be expressed in terms of the specific angular momentum $h=\sqrt{GMr_a(1-e)}$ and the specific energy $k=-GM/(2r_a)$ (with $r_a$ is the semi-major axis). Following the derivation of \citet{wangOrbitalEvolutionTidally2024}, the phase-dependent changes are  
% \begin{equation}
% \begin{aligned}
%     \frac{\dot{h}}{h}&=\frac{1}{\tau_\mathrm{F}}\left(\frac{1}{\sqrt{1+e\cos f}}-1\right)\\
%     \frac{\dot{k}}{k}&=\frac{2}{\tau_{\mathrm{F}}}\left(\frac{\left(1+2 e \cos f+e^{2}\right)-(1+e \cos f)^{3 / 2}}{1-e^{2}}\right)
% \end{aligned}
%     \label{eq:hhkk}
% \end{equation}
% where $f$ is the true anomaly. Then the secular evolution of orbital parameters:
% \begin{equation}
%     \begin{aligned}
%         \left\langle\frac{\dot a}{a}\right\rangle _{\text{drag}}&=-\left\langle\frac{\dot k}{k}\right\rangle\\
%         \left\langle\frac{\dot e}{e}\right\rangle _{\text{drag}}&=-\frac{1-e^2}{e}\left( \left\langle\frac{\dot h}{h}\right\rangle+\frac{1}{2}\left\langle\frac{\dot k}{k}\right\rangle\right)
%     \end{aligned}
%     \label{eq:drag_ae}
% \end{equation}
% and the change in penetration factor $\beta$ is
% \begin{equation}
%     \left\langle\frac{\dot{\beta}}{\beta}\right\rangle_{\text {drag}}=-\frac{1+e}{e}\left\langle\frac{\dot{h}}{h}\right\rangle-\frac{1-e}{2 e}\left\langle\frac{\dot{k}}{k}\right\rangle
%     \label{eq:betadot_disk}
% \end{equation}

Then the secular evolution of orbital parameters:
\begin{equation}
    \begin{aligned}
        \left\langle\frac{\dot a}{a}\right\rangle _{\text{drag}}\propto -\frac{1}{\tau_F}, &\quad \left\langle\frac{\dot e}{e}\right\rangle _{\text{drag}}\propto -\frac{1-e^2}{2e}\frac{1}{\tau_F}\\
        \left\langle\frac{\dot{\beta}}{\beta}\right\rangle_{\text {drag}} & \propto-\frac{1-e}{2e} \frac{1}{\tau_F}
    \end{aligned}
    \label{eq:drag_dot}
\end{equation}
% and the change in penetration factor $\beta$ is
% \begin{equation}
%     \left\langle\frac{\dot{\beta}}{\beta}\right\rangle_{\text {drag}}=-\frac{1+e}{e}\left\langle\frac{\dot{h}}{h}\right\rangle-\frac{1-e}{2 e}\left\langle\frac{\dot{k}}{k}\right\rangle
%     \label{eq:betadot_disk}
% \end{equation}

\par \citet{wangOrbitalEvolutionTidally2024} discusses      that 
% \par The signs of Eqs. (\ref{eq:drag_ae},\ref{eq:betadot_disk}) imply that 
the drag force enhances GW emission ($\langle\dot a\rangle<0$ and $\langle\dot e\rangle<0$) but hinders MT ($\langle \dot \beta\rangle<0$). Physically, the drag shrinks the orbital and drives circularization by enhancing the rate of orbital energy loss, 
thereby reinforcing the GW emission, yet it simultaneously moves the orbit away from a periapsis, which suppresses MT. For our fiducial parameters (GSN 069 with $\beta_\text{initial}=0.375,\mu_\text{WD}=0.3M_\odot$ indicated by Fig. \ref{fig:timescale_beta_gsn069}), Eq. \eqref{eq:drag_dot} yields $\langle{\dot \beta/\beta}\rangle_{\text{disk}}\sim \mathcal O(10^{-4}) \text{ yr}^{-1}$, dominating over purely GW-driven $\langle{\dot \beta/\beta}\rangle_{\text{GW}}\sim \mathcal O(10^{-5}) \text{ yr}^{-1}$. Hence, the drag force tends to delay the onset of the MT-dominated phase period.   

\par If the drag effect is not considered, the cancellation between GW-driven orbital shrinking and MT-driven orbital widening arises naturally at the intermediate stage of the secular orbital evolution. This cancellation can explain the observed disappearance of the recurrence time decay, as shown in Sect. \ref{sec:orb-evo-stripping}. However, including additional drag torques from the accretion flow may dominate the orbital evolution, which is similar to the orbiter-disk collision scenario, resulting in the absence of the cancellation. Thus, involving the drag force may introduce a parameter-tuning problem for producing the observed orbital evolution.

\par From the standpoint of detectability, the environmental effects modify GW emission relative to vacuum templates and should be included when modelling the secular dynamics and the GW waveform (see Sect. \ref{sec:environment}).
In this case, the environmental effect is no longer neglected in the secular orbital evolution and GW waveform calculation. However, even neglecting all the environmental effects, the GSN 069 with baseline initial parameters mentioned above yields an SNR $\ll 1$ after 15 yr of GW-only orbital evolution and a 4-yr LISA observation; eRO-QPE3 and eRO-QPE4 yield even smaller SNRs due to their larger orbits. Therefore, though disk drag is important for orbital evolution and GW waveform, it does not alter our qualitative conclusion about GW detectability for these sources.  

\par Notably, we do not include the tidal heating effect induced by repeated periapsis passages, which can significantly modify the structure of the WD. Dynamical tides can inject energy into the WD, dissipating much larger energy than the binding energy, which leads to a faster orbital evolution \citep{yaoMassTransferTidally2025}. In addition, the tidal heating inflates the WD radius, which enhances the MT. Therefore, this effect will shorten the $\tau_\text{GW-d}$. 

\subsubsection{For the orbiter-disk collision scenario: collision in region A} \label{sec:regionA}

%%%%%%%%
%% 放到appendix或许会更好
%%%%%%%%
\par In our baseline analysis, we assume that all collisions occur in region B of the standard disk, i.e., at radius $R$ exterior to the A–B transition radius $R_\text{AB}$ set by Eq. \eqref{eq:rab}. The scaling in Eq. \eqref{eq:rab}, $R_\text{AB}\approx M^{2/21} \dot m^{16/21}$, implies that increasing either the central black-hole mass $M$ or the accretion rate $\dot m$ moves the transition radius outward. In systems with sufficiently large $M$ and $\dot m$, the inferred collision radius $R$ can therefore lie inside $R_{\rm AB}$, placing the interaction in region A. For example, if we take the central mass of GSN 069 to be $M=10^6M_\odot$ and the accretion rate to be $\dot m = 0.1$, the $R_\text{AB}$ should be $\sim 250 R_g$, which is larger than $R\sim 160 R_g$. Therefore, we explore the case of collisions within region A.

\par According to Eq. \eqref{eq:pdot_sbh} and the equations in Appendix \ref{app:disk_structure}, the orbital decay rate in region A should be:
\begin{equation}
\begin{aligned}
     \dot P_\text{orb,A}&\simeq -1.0\times 10^{-5} \text{ s s}^{-1} \left(\frac{0.1}{\alpha}\right)\left(\frac{P}{5\text{h}}\right)^{7/3}  \\
     &\left(\frac{\ln \Lambda}{10}\right) \left(\frac{0.1}{\dot m}\right) \left(\frac{\mu}{100M_\odot}\right) \\
     &\left(\frac{\sin\theta_\text{od}}{0.02}\right)^{-3}\left(\frac{M}{10^6 M_\odot}\right)^{-7/3}
     \label{eq:pdot-regionA}
\end{aligned}
\end{equation}

\par We fixed $\dot m = 0.1$, while taking both the central masses of GSN 069 and eRO-QPE2 to $10^6M_\odot$ (while retaining the fiducial masses for eRO-QPE3 and eRO-QPE4 so that $R<R_{\rm AB}$ holds), and keep all remaining parameters as in the baseline model. Moreover, the misalignment angle $\theta_{\text{od}}$ must be significantly small to match observed orbital decay, so we set it to $\sin\theta_{\rm od}=0.02$.

\par For the sBH-disk collision scenario, GSN 069 remains undetectable with $\rm SNR \ll 1$. The eRO-QPE2 yields a smaller SNR of 3.1 (calculated with \texttt{FEW}), with a mean orbital decay rate of $\langle\dot P_\text{orb}\rangle \simeq 1.6\times 10^{-6} \text{s s}^{-1}$. Hence, eRO-QPE2 would be undetectable if the collision occurs in region~A at higher $M$. For eRO-QPE3 and eRO-QPE4, the results have only gentle differences with those in the main analysis, since the observational constraint $\langle\dot P_\text{orb}\rangle \simeq \mathcal O(10^{-6}) \text{s s}^{-1}$ limits the parameter shifts. Therefore, we omit discussions of both sources. 
\par For the iBH-disk collision scenario, we choose $\sin\theta_{\rm od}=0.04$ and $\sin\theta_{\rm od}=0.1$ for GSN 069 and eRO-QPE2 to ensure $\langle\dot P_\text{orb}\rangle \simeq \mathcal O(10^{-6}) \text{s s}^{-1}$. The resulting SNRs are ${\rm SNR}=2.1$ for GSN 069 and ${\rm SNR}=42.8$ for eRO-QPE2, indicating that GSN 069 would remain below detectability while eRO-QPE2 would be a confident detection under the same observing setup.
\par In summary, the GW detectability of the collision is sensitive to the central mass. A more precise constraint on the central mass benefits the prediction of GW detectability. 

\subsubsection{For the orbiter-disk collision scenario: warp–driven alignment of the disk}\label{sec:aligment}
\par In the orbiter-disk collision model, the secular evolution of the orbital period is attributed to the Lense-Thirring precession of the disk and orbiter. Internal stresses within a viscous disk act to communicate and damp warps; as a result, the inner disk tends to align with the black-hole equatorial plane, which is called Bardeen–Petterson effect \citep{bardeenLenseThirringEffectAccretion1975}. Once the disk has aligned, the relative misalignment angle between the orbiter and the disk $\theta_\text{od}$ keeps constant, and the evolution of orbital decay stops.

\par There are two types of viscosity in the standard disk:
\begin{itemize}
    \item viscosity from the co-plane shear between neighboring annuli due to differential rotation, denoted as $\nu_1$.
    \item viscosity from the vertical shear (warp) with the disk between neighboring annuli due to disk twist, denoted as $\nu_2$.
\end{itemize}
\par With small warp, the definition of these two viscosities is given by \cite{ogilvieNonlinearFluidDynamics1999}:
\begin{equation}
    \begin{aligned}
\nu_1&=\alpha c_s H=\alpha H^2\Omega_K\\
\nu_2&=\left[\frac{2(1+7\alpha^2)}{\alpha^2(4+\alpha^2)}\right]\nu_1 \simeq \frac{\nu_1}{2\alpha^2}  \quad \text{  for  } \alpha^2 <<1
\end{aligned}
\label{eq:viscosity}
\end{equation}
where $c_s$ is the sound speed of the disk and $\Omega_K=2\pi/P_{\text{orb}}$ is the orbital angular frequency. 
%% linear
\par  In the diffusive warp regime of the disk (thin disk with $\alpha >H/R$ and small warp) relevant to QPE systems, local damping of a warp proceeds on the warp–diffusion time,
\begin{equation}
    t_\text{align}\sim t_\text{diff}=R^2/\nu_2
\end{equation}
under linear approximation. The vertical viscosity damps the local Lense-Thirring precession. Substituting Eq. \eqref{eq:hr_zoneB} \eqref{eq:hr_zoneA} and \eqref{eq:viscosity} to the expression, the warp-diffusion time in region A and region B should be:

\begin{equation}
\begin{aligned}
    t_\text{align,A}&=14\text{ d }\left(\frac{\alpha}{0.1}\right)\left(\frac{P}{5\text{h}}\right)^{7/3}\left(\frac{M}{10^6M_\odot}\right)^{-4/3}\left(\frac{\dot m}{0.1}\right)^{-2}\\
    t_\text{align,B}&=1.5\text{ yr }\left(\frac{\alpha}{0.1}\right)\left(\frac{P}{5\text{h}}\right)^{14/15}\\& ~~~~~~~~~~~~~\left(\frac{M}{10^5M_\odot}\right)^{4/15}\left(\frac{\dot m}{0.05}\right)^{-2/5}
\end{aligned}
    \label{eq:tdiff_regionA}
\end{equation}

\par For QPE-like parameters, both $t_\text{align,A}$ and $t_\text{align,B}$ are typically much shorter than the multi-year evolution inferred from the observed recurrence-time trends. Moreover, the parameter dependencies expose a tension: increasing $t_\text{align,A}$ favors smaller $M$ and $\dot m$, both both choices shrink region A and can move the collision radius out of it; in region B, $t_\text{align,B}$ grows with larger $M$ and smaller $\dot m$, yet the power index of them are small, yielding only modest lengthening.

\par This discrepancy suggests that linear, diffusive warp theory is insufficient to explain delayed alignment in QPE disks. Non-linear warp theory may be needed to introduce a stronger warp, leading to a slower propagation\citep{ogilvieNonlinearFluidDynamics1999}. 
Alternatively, the orbiter-disk interaction reduces the radial communication between annuals or even breaks the disk, increasing the warp-diffusive timescale \citep{nixonBrokenDiscsWarp2012}. Moreover, anisotropic viscosity driven by magneto-rotational instability stresses that deviate from the isotropic case \citep{nixonAccretionDiscViscosity2015}, which may slow down the warp propagation. These effects significantly complicate the analysis, which is out of the scope of this paper. Considering these effects benefit for comprehensively modeling QPE and better understanding the properties of the accretion disk.

\subsubsection{Environmental effects on GW detection}\label{sec:environment}
In our fiducial analysis of both scenarios, we neglect environmental perturbations to the GW waveform and employ vacuum templates. Gas–related processes can dephase the signal relative to vacuum evolution, introducing a small mismatch, through hydrodynamic drag, dynamical friction, planetary-type migration torques, and other disk–orbiter couplings \citep{barausseInfluenceHydrodynamicDrag2008,yunesImprintAccretionDiskInduced2011,barausseCanEnvironmentalEffects2014,barausseEnvironmentalEffectsGravitationalwave2015,cardosoConstraintsAstrophysicalEnvironment2020,speriProbingAccretionPhysics2023}. However, in the LISA band (mHz), the environmental effects are negligible in most cases, with GW phase shift less than $10^{-2} \text{ rad/yr}$ for the EMRI/IMRI systems \citep{barausseCanEnvironmentalEffects2014,barausseEnvironmentalEffectsGravitationalwave2015}. The environmental effect becomes more relevant when geometrically-thin, radiatively efficient accretion disks are present. 

Observationally, several QPE sources follow after tidal disruption events (TDEs) in years \citep{miniuttiNinehourXrayQuasiperiodic2019,nichollQuasiperiodicXrayEruptions2024,chakrabortyDiscoveryQuasiperiodicEruptions2025}. Motivated by this, many implementations of the orbiter–disk collision scenario (including the one used in this work) assume a pre-existing accretion TDE disk around the SMBH \citep{linialEMRITDEQPE2023,linialCoupledDiskstarEvolution2024,zhouProbingOrbitsStellar2024a}. The TDE disk is geometrically thick and radiatively inefficient at early times, whereas at later times it transitions to a geometrically thin, radiatively efficient disk as the accretion rate declines \citep{bambiHandbookXrayGammaray2024}, when the QPE happened. Environmental dephasing is therefore more pronounced with the onset of QPEs. However, a recent work by \citet{luiGravitationalWaveSignatures2025} shows that discrete orbiter–disk impacts can \emph{enhance} GW emission by injecting additional high-frequency harmonics, while leaving the low-frequency content essentially unchanged for a low-eccentricity orbit ($e\sim 0$). Due to its low amplitude, the introduced harmonics may not impact the parameter estimation of the GW in the orbiter-disk collision scenario. Thus, adopting a vacuum waveform in GW parameter estimation may not introduce significant bias.
\par Similar considerations apply in the stripping scenario if a residual TDE disk is present. By contrast, the debris periodically stripped from the WD is expected to circularize to a thick mini-disk, similar to the early stage of a TDE disk \citep{narayanAdvectiondominatedAccretionUnderfed1995,yuanHotAccretionFlows2014}, further suppressing GW dephasing. 

\subsection{Potential detections of bright siren}\label{sec:discussion-joint}

\par Recent surveys indicate a promising discovery rate for QPEs, with forecasts of $\sim$1–4 new sources per year \citep{arcodiaXrayQuasiperiodicEruptions2021,arcodiaCosmicHideSeek2024}. Thus, it's possible to discover additional \textit{regular} QPEs in the next few years. Regular sources enable modeling of orbital evolution and, in turn, predictions for GW detectability. The currently known sample (currently two regular, two irregular, and five unresolved sources) suggests that partial sources exhibit sufficiently stable timing for such modeling. In particular, the discovery of a regular QPE with a compact orbit (akin to eRO-QPE2) would offer a realistic prospect for detection with LISA in the 2030s

\par Related nuclear transients may also serve as bright sirens. Several less intense eruptions around SMBHs, called quasi-periodic oscillations (QPOs), have been discovered recently \citep{gierlinskiPeriodicity1Hour2008,mastersonMillihertzOscillationsInnermost2025}. QPOs show some similarities to QPEs, such as flaring properties, the properties of the host galaxies, and SMBHs. However, the QPO periods (from about 7 to 60 minutes) are much shorter than those of QPEs. Some of these QPOs are also probably produced by EMRI companions around SMBHs \citep{mastersonMillihertzOscillationsInnermost2025}, thereby expanding the pool of prospective EMRI bright sirens.  

\par Two concrete examples illustrate possibilities of QPO bright sirens. A QPO source RE J1034+396 shows an about 250-second decrease in recurrence time between 2007 and 2020  \citep{gierlinskiPeriodicity1Hour2008,jinReobservingNLS1Galaxy2020}, indicating a long-term orbital evolution, despite mid-term irregularity. Modeling this system as an orbiter–disk collision powered by a compact object of $\sim 46 M_\odot$ yields a GW signal with an SNR $\sim 13.8$ in the 2030s after an orbital evolution started from 2007 \citep{kejriwalRepeatingNuclearTransients2024}. By contrast, another QPO source 1ES 1927+654 demonstrates much faster evolution, with recurrence time decay from an 18 min ($\sim \mathcal O(0.1)\text{ mHz}$) period down to 7 min ($\mathcal O(1)\text{ mHz}$) in about two years \citep{mastersonMillihertzOscillationsInnermost2025}. If interpreted as stable mass transfer from a companion, such a configuration could be detected by LISA with an SNR $\sim 10$, even without long-term orbital evolution \citep{karaSupermassiveBlackHoles2025}. However, this system will coalesce before 2030 if the current rapid recurrence time decay persists.

\par With anticipated discovery rates and the growing census of QPEs and QPOs, we expect a handful of regular, short-period systems to enter the LISA band with measurable chirps in the 2030s. Long-baseline monitoring is essential to identify sources with stable timing and predict their GW detectability.

\subsection{Detectability enhancement due to mission lifetime and detector networks}
\par
\begin{figure}[htbp]
    \centering
    \includegraphics[width=\linewidth]{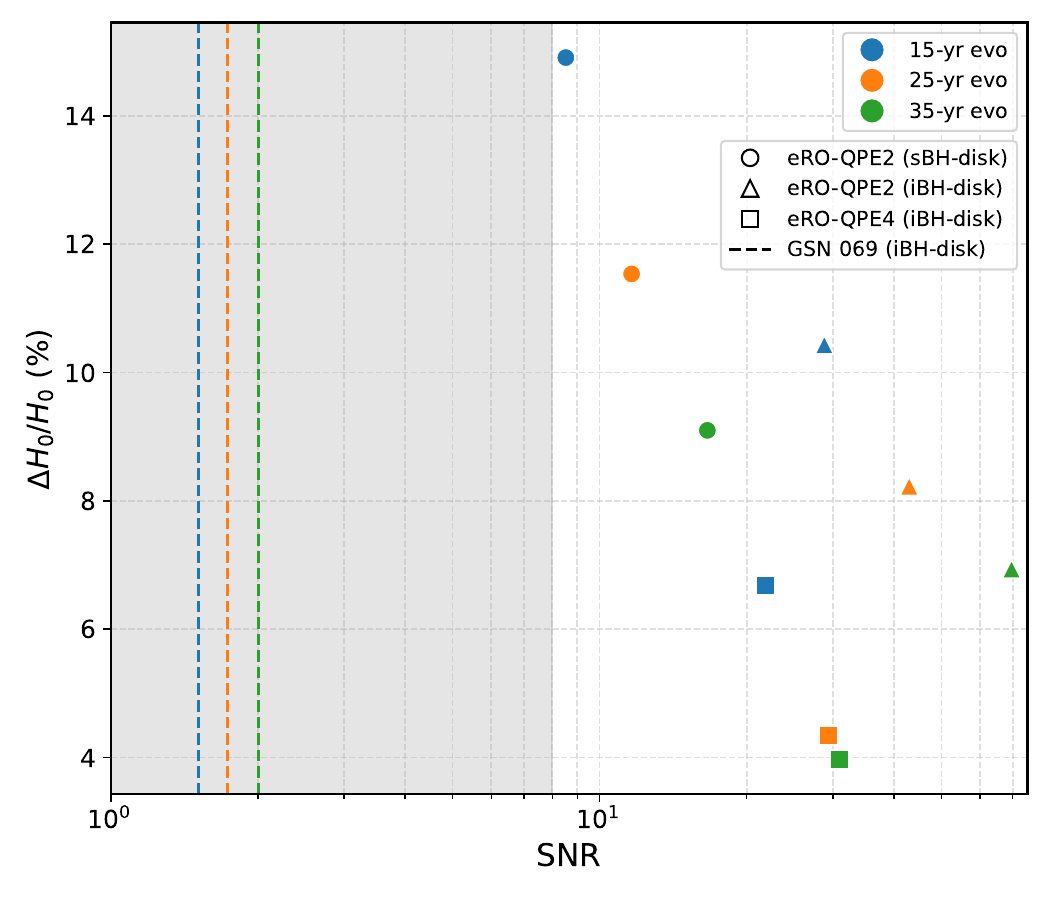}
    \caption{The SNR vs the fractional $H_0$ uncertainty, $\Delta H_0/H_0$ for sources in our fiducial analysis with 4-yr LISA observation. Colors encode the elapsed secular orbital evolution (blue: 15 yr; orange: 25 yr; green: 35 yr). Symbols denote systems and scenarios: circles—eRO-QPE2 (sBH–disk); triangles—eRO-QPE2 (iBH–disk); squares—eRO-QPE4 (iBH–disk).. The dashed lines represent GSN 069 in the iBH-disk scenario. The gray shaded band marks the SNR threshold of $\leq 8$, i.e., below our detection threshold. As the evolution time increases, all sources move toward higher SNR and yield progressively tighter constraints on $H_0$. We plot the sources with SNRs over 1 in the figure.}
    \label{fig:snr-h0}
\end{figure}

 The LISA mission is planned for launch in 2035 and is designed to operate for 4.5 years, with consumables allocated to support a potential extension to 10 years \citep{colpiLISADefinitionStudy2024}. A longer mission allows QPE systems to undergo additional secular orbital evolution before and during the observation window, thereby strengthening their GW signals. As shown in Fig. \ref{fig:snr-h0}, we calculate the SNR of detectable sources with assumed evolution time (15, 25, and 35 years) and their corresponding constraint on $H_0$. With longer evolution time, all sources yield larger SNR and tighter constraints on $H_0$. 
\par Despite this trend, no additional sources cross our detection threshold (SNR $\geq 8$) with longer evolution up to 35 years. For the orbiter-disk collision scenario, only two sources are detectable, namely eRO-QPE2 and eRO-QPE4. eRO-QPE2 in the sBH-disk scenario yields minimal SNR and the loosest constraint on $H_0$ (with SNR$\sim 8.5-16.6$ and $\Delta H_0/H_0\sim 9.1-14.9\%$). eRO-QPE2 in the iBH-disk scenario yields the largest SNR ($\sim 28.8-69.7$), but eRO-QPE4 in the iBH-disk scenario constrains the $H_0$ best ($\sim 4.0-6.7\%$), due to its large luminosity distance, leading to less uncertainty of peculiar velocity. For the stripping scenario, GSN 069 is the source with the largest SNR, though the SNRs are still under the threshold with 35-yr evolution.   

\par It's necessary to point out that the GW detectability and constraints on $H_0$ are solely derived from the LISA observation. Additional space-based detectors are expected to operate in the 2030s— Taiji \citep{huTaijiProgramSpace2017} and Tianqin \citep{luoTianQinSpaceborneGravitational2016}. These detectors exhibit similar sensitivity to LISA, leading to a similar SNR for a single source. If these space-based detectors have a significant time overlap in the 2030s, the detector network could lead to enhanced SNRs for detected sources by a factor $\sim 2$, according to $\rho=\sqrt{ \rho_{\text{LISA}}^2+\rho_{\text{Taiji}}^2+\rho_{\text{Tianqin}}^2}$. 
Higher SNRs, together with multi-detector timing, can improve sky localization and reduce uncertainties in luminosity distances \citep{kyutokuGravitationalwaveCosmographyLISA2017,zhanBrightSirenElectromagnetic2025}, thereby tightening constraints on $H_0$. 

\par Looking further ahead, next-generation space-based detectors, Deci-Hertz Interferometer Gravitational wave Observatory (DECIGO)\citep{kawamuraJapaneseSpaceGravitational2011} and Big Bang Observer (BBO)\citep{harryLaserInterferometryBig2006},  are scheduled to be operational in the next few decades. These detectors would further enhance the detectability of the GW sources. Nevertheless, continued improvements in detector capabilities, together with a clearer mapping between QPE/QPO phenomenology and EMRI dynamics, will sharpen GW-based measurements of the cosmic expansion and other cosmological parameters.

\section{Conclusion}\label{sec:conclusion}
\par QPE systems are expected to be associated with EMRIs or IMRIs, which are the main target of the space-borne GW detector, such as LISA. We have assessed the GW detectability of known QPEs by modeling their secular orbital evolution in two physical frameworks: (i) a stripping scenario, in which eruptions are powered by periodic MT from the orbiter to the central SMBH; and (ii) an orbiter–disk collision scenario—considering both sBH–disk and iBH–disk collisions—in which eruptions are triggered by recurring impacts between the orbiter and an accretion disk. Partial QPEs exhibit a regular decay of recurrence time, which can subsequently flatten. In our interpretation, the regular decay comes from orbital energy loss due to MT in the stripping scenario and orbiter-disk interaction in the collision scenario. By contrast, the decay-flattening behavior is explained by a transition from GW-dominated to MT-dominated phase in the stripping scenario, and the coupled precession of the orbiter and disk in the collision scenario.

We calculate the orbital decay and waveform by \texttt{FEW} for the stripping scenario. We calculate the orbital decay by mean decay rate and calculate the waveform by \texttt{FEW} or \texttt{IMRphenomD} depending on the mass ratio of the system. The environmental effect in the GW waveform calculation is not considered in this work.

\par The main results are summarized as follows:
\begin{enumerate}
    \item \textit{Stripping scenario.} None of the current QPE sources are detectable by LISA. The GW-dominated period $\tau_\text{GW-d}$, constrained by observations, is too short for secular evolution for GSN 069, eRO-QPE2, and eRO-QPE4. Their orbital evolution would not be dominated by GW in the 2030s, leading to negative detectability. For the eRO-QPE3, the $\tau_\text{GW-d}$ allows it to decay for 15 years; however, it yields an SNR $\ll 1$ in a 4-year observation.  
    
    \item \textit{sBH–disk collision} (with $\mu=100M_\odot$). For eRO-QPE2, we find an SNR of 8.5 over a 4-yr observation, corresponding to a mean orbital decay rate $\langle \dot P_\text{orb}\rangle\simeq -6.0\times 10^{-6} \text{s s}^{-1}$. Such a detection would constrain the $H_0$ with a fractional uncertainty of $\Delta H_0/H_0 \simeq 14.9\%$.

    \item \textit{iBH–disk collision} (with $\mu=10^{3}M_\odot$). eRO-QPE2 and eRO-QPE4 yield SNRs of 28.8 and 21.8, respectively, enabling $\Delta H_0/H_0 \simeq 10.4\%$ (eRO-QPE2) and $6.7\%$ (eRO-QPE4), under mean orbital decay rates of $-7.3\times 10^{-6}\text{s s}^{-1}$ and $-6.3\times 10^{-7}\text{s s}^{-1}$.
    
    \item A longer operating lifetime strengthens all constraints. Extending the evolution time from 15 to 35 years, eRO-QPE2 in the sBH-disk case remains the weakest of the collision scenarios, with SNR$\sim 8.5-16.6$ and $\Delta H_0/H_0\sim 9.1-14.9\%$. In the iBH–disk case, eRO-QPE2 in the iBH-disk scenario yields the largest SNR ($\sim 28.8-69.7$), but eRO-QPE4 in the iBH-disk scenario constrains the $H_0$ best ($\sim 4.0-6.7\%$), due to its large luminosity distance, leading to less uncertainty of peculiar velocity.
\end{enumerate}

In summary, QPE systems evolving under the orbiter–disk collision scenario—particularly those with an iBH companion—emerge as promising space-based GW sources and viable \emph{bright sirens} for precision cosmology in the 2030s. Incorporating environmental effects, eccentricity–harmonic content, and disk–precession dynamics self-consistently into the waveform will be the essential next steps to robustly forecast detectability and to refine the resulting constraints on $H_0$.
%\section*{Acknowledge}


%\begin{appendices}

% \end{appendices}

%\begin{addendum}
\section*{Acknowledgements}
This work was supported by the National Natural Science Foundation of China (grant Nos. 12494575 and 12273009).

\appendix
\section{The recurrence time and orbital period} \label{app:Trec_Porb}
For the stripping scenario, we assume a WD companion on an extremely eccentric orbit (eccentricity $e\gtrsim 0.95$). If one eruption is launched per orbit, the recurrence time decay rate $\dot T_\text{rec}$ is naturally the sum of the orbital period decay rate $\dot P_\text{orb}$ and the decay of characteristic accretion time $\dot t_\text{acc}$ for the stripped debris, including fallback and circularization,
\begin{equation}
    \dot T_\text{rec}\simeq \dot P_\text{orb}+\dot t_\text{acc}
\end{equation}
Since $t_\text{acc}$ is stable for each stripping \citep{shenFastUltraluminousXRay2019}, we assume $\dot t_\text{acc}\sim 0$ and therefore adopt $\dot T_\text{rec}\simeq \dot P_\text{orb}$ for this scenario.

\par By contrast, we assume that the inspiral in a circular orbit ($e=0$) for the orbiter-disk collision scenario. Eruptions are triggered when the orbiter intersects the precessing disk; hence, $\dot T_\text{rec}$ is set by the twice-per-orbit crossing frequency and the local Lense–Thirring precession of the disk at the collision radius
\begin{equation}
    \dot T_\text{rec}\simeq \frac{2\pi (2\dot \Omega_\text{orb}-\dot \Omega_\text{LT})}{(2\Omega_\text{orb}-\Omega_\text{LT})^2}
    \label{eq:trec}
\end{equation}
where $\Omega_\text{orb}, \Omega_\text{LT}$ are the orbital frequency and Lense-Thirring precession frequency of the disk, respectively, and the right-hand side is derived from the differential rotation period between the orbiter and the disk. The $\Omega_\text{LT}$ is stable and much lower than $\Omega_\text{orb}$ (see Sect. \ref{sec:orb-evo-collision} for detail), leading to $\dot T_\text{rec}\simeq \dot P_\text{orb}/2$. 

\par Thus, in both scenarios, we assume that the orbital evolution tracks the recurrence time decay, i.e., $\dot T_\text{rec}\simeq \dot P_\text{orb}$ for the stripping scenario and $\dot T_\text{rec}\simeq \dot P_\text{orb}/2$ for the orbiter-disk scenario. Observation indicates $\dot T_\text{rec}$ has a trend to vanish. We should match the evolution of $\dot T_\text{rec}$ with the self-consistent evolution of $\dot P_\text{orb}$.

\section{Disk structure} \label{app:disk_structure}
\par A standard $\alpha$-accretion disk can be separated into three radial regions \citep{shakuraBlackHolesBinary1973}: (A) radiation-pressure-dominated inner region; (B) a gas–pressure–dominated region with Thomson (electron–scattering) opacity; and (C) a gas–pressure–dominated region with Kramers (free–free/bound–free) opacity. Since the geometry and the properties in each region are different, the local drag and the orbital decay depend critically on where the collision occurs. The transition radius $R_\text{AB}$ between region A and region B can be simply estimated \citep{lipunovaStandardModelDisc2018}:
\begin{equation}
    {R_\text{AB}}\sim 118 R_g\left(\frac{\alpha}{0.1}\right)^{2/21} \left(\frac{M}{10^5 M_\odot}\right)^{2/21}\left(\frac{\dot m}{0.05}\right)^{16/21}
    \label{eq:rab}
\end{equation}
where $R_g\equiv GM/c^2$ is the gravitational radii, and $\alpha$ is the viscosity parameter. 

\par For example, for the two regular QPEs—GSN 069 ($M=3\times 10^5M_\odot, P_\text{orb}=18\text{ h}$) and eRO-QPE2 ($M=10^5M_\odot,P_\text{orb}=5\text{ h}$)—the inferred orbital radius is $R\sim 300 R_g$, indicating that $R>R_\text{AB}$. Thus, we assume that all the orbiter–disk impacts occur in region B of the disk for all sources. 

\par Except for region B, we also calculate the cases that collision occurs in region A. In region A, where radiation pressure dominates over the gas pressure of the disk, the $H/R$ can be expressed by \citep{frankAccretionPowerAstrophysics2002}:   
\begin{equation}
    \left(\frac{H}{R}\right)_A=\frac{3}{2\eta} \frac{R_g}{R} \dot m\simeq 1.5\times 10^{-2} \left(\frac{\dot m}{0.1}\right)\left(\frac{100R_g}{R}\right)
    \label{eq:hr_zoneA}
\end{equation}
where we set the radiation efficiency to a typical value, $\eta=0.1$. In addition, the surface density $\Sigma_A$ in region A of the disk can be expressed by 
\begin{equation}
\begin{aligned}
    \Sigma_A&\simeq \frac{\dot M}{3\pi \nu_1}\\
    & \simeq 1.7\times 10^4 \mathrm{g ~cm^{-2}} \left( \frac{0.1}{\alpha} \right)\left( \frac{0.1}{\dot m} \right) \left( \frac{R}{100R_g} \right)^{3/2}
\end{aligned}
    \label{eq:sigma_zoneA}
\end{equation}
where $\nu_1$ is the kinetic viscosity of the disk \citep{shakuraBlackHolesBinary1973}.

\section{The challenges and benefits of the BH-disk collision scenario}\label{app:bh-disk_challenge}  
\par Notably, the BH-disk collision scenario faces two principal challenges: the energy budget may not be enough for the a sBH companion \citep{linialEMRITDEQPE2023,lamBlackHoleaccretionDisk2025}; the event rate may be too low for a iBH \citep{linialEMRITDEQPE2023}. A full reconciliation of these issues lies beyond the scope of this work. We address a portion of the challenges and highlight conditions under which the BH-disk collision scenario benefits. 
\par A two-dimensional hydrodynamic simulation of BH-disk collision suggests that a QPE system requires a BH orbiter with a mass of $10^4-10^5 M_\odot$ for an orthogonal collision case (with the orbit–disk misalignment angle $\theta_\text{od} \pi/2$), which is consistent with \citep{linialEMRITDEQPE2023}. However, the effective cross section for the BH orbiter increases with $\theta_\text{od}$, due to the decrease of the relative velocity between the orbiter and the disk. Consequently, small tilts can offset the low value of the effective radius, in turn yielding enough energy. A second, qualitative discriminator is the morphology of the ejecta. The BH-disk collision tends to produce more symmetric outflows in both sides of the disk \citep{ivanovHydrodynamicsBlackHoleAccretion1998,lamBlackHoleaccretionDisk2025}, compared to asymmetric ejecta from the star-disk collision \citep{huangMultibandEmissionStarDisk2025}. Finally, constraints that the companion not be tidally disrupted by the central SMBH favor a compact object over a main-sequence star, indirectly supporting a BH companion in this framework \citep{guoTestingStardiskCollision2025}.
% \section{Author Contributions}
% F.Y. W. conceived the project. Y.J. Z. led the analysis, interpretation, simulation, and drafted the manuscript. F.Y. W. analyzed the models and edited the manuscript. D. W. provided a discussion on the mass transfer in the stripping model. S.X. Y. discussed the QPE models. All authors contributed to the scientific interpretation of the results.
% \section{Competing Interests} The authors declare no competing interests.

% \section{Correspondence} Correspondence and requests for materials should be addressed to \\
% F.Y. W. (fayinwang@nju.edu.cn).

% \section{Data availability}
% The data and codes for reproducing the results of this work are available from the corresponding author upon reasonable request.
% \appendix

% \bibliography{ms}% common bib file
\bibliography{Cosmology}% common bib file

\end{document}